\documentclass[reprint,amsmath,amssymb,aps,prx,
  nofootinbib,longbibliography,floatfix]{revtex4-2}

\usepackage{graphicx}
\usepackage{bm}
\usepackage{hyperref}
\hypersetup{hidelinks}
\usepackage{xcolor}

\graphicspath{{figures/}}

\newcommand{\pversion}{\texttt{cWB-space}}
\newcommand{\Achan}{A}
\newcommand{\Echan}{E}
\newcommand{\Tchan}{T}
\newcommand{\XYZ}{X,Y,Z}
\newcommand{\AET}{A,E,T}

\begin{document}

\title{cWB-space: A time-frequency unmodelled transient-search pipeline for LISA}

\author{Shubhanshu Tiwari}
\email[Corresponding author: ]{shubhanshu.tiwari@eaps.ethz.ch}
\affiliation{%
Department of Earth and Planetary Sciences, ETH Zurich,\\
Sonneggstrasse 5, 8092 Z\"urich, Switzerland
}
\author{Yumeng Xu}
\affiliation{Departament de F\'isica, Universitat de les Illes Balears, IAC3 --
IEEC, Crta. Valldemossa km 7.5, E-07122 Palma, Spain}
\author{Giovanni A Prodi}
\affiliation{Department of Physics, University of Trento, Trento, Italy}
\affiliation{ INFN-TIFPA, Trento Institute for Fundamental Physics and
Applications, Trento, Italy}

\date{\today}

\begin{abstract}
The Laser Interferometer Space Antenna (LISA) will observe
massive-black-hole mergers and other transient phenomena alongside many
long-lived gravitational-wave signals. Some transients may be
incompletely modeled or have waveforms that have not been anticipated.
We present \texttt{cWB-space}, a pipeline that combines an unmodeled
transient search, waveform reconstruction, and a time-frequency glitch
classifier. The search identifies concentrations of signal power in time
and frequency and reconstructs candidate waveforms without requiring a
source template. The glitch classifier compares how candidates appear
in different combinations of LISA's laser measurements, with the
background estimated from the observed data. We test the pipeline with
simulated black-hole binaries with aligned and precessing spins, burst
signals with different shapes, and instrumental glitches. The response
comparisons help assess a candidate's astrophysical or instrumental
origin, while also revealing correlated glitches that can resemble
gravitational waves. In the Sangria simulated LISA data, the search
recovers all six injected massive-black-hole binaries as its six
highest-ranked candidates. Their local reconstructions recover the main
features of the injected waveforms over the intervals studied. These
results demonstrate how an unmodeled search can connect transient
identification with investigation of a candidate's origin and waveform
recovery.
\end{abstract}

\maketitle

\section{Introduction}
\label{sec:introduction}

The Laser Interferometer Space Antenna (LISA) will observe massive-black-hole
mergers and other transient phenomena alongside a large population of
long-lived gravitational-wave
sources~\cite{LISADefinitionStudy2024,NASALISAMissionReference}. These
observations will probe the dynamics of compact objects and may reveal
sources that have not previously been observed. Accurate waveform models
will be central to this effort, but searches and waveform reconstruction
must also accommodate signals whose detailed waveforms are uncertain.
Complex source dynamics can make waveform modeling challenging, while
unexpected sources may produce signals outside the waveform families
considered in an analysis. This motivates unmodeled searches, which
identify concentrations of signal power without requiring a detailed
source template. Following this power in time and frequency can support
both the discovery of unexpected phenomena and the investigation of known
sources whose waveforms are incompletely modeled. Here, ``transient''
denotes a signal whose power is predominantly concentrated within a time
interval much shorter than the observation, even when the source itself
evolves over a much longer timescale.

Ground-based observations have established unmodeled searches as an
important part of gravitational-wave astronomy. The coherent WaveBurst
(cWB) pipeline~\cite{Klimenko2016TransientMethod} has repeatedly detected
binary-black-hole mergers and contributed search results to successive
Gravitational-Wave Transient Catalogs, alongside searches based on waveform
templates~\cite{GWTC2,GWTC3,GWTC4}. Two prominent examples are GW150914,
initially identified by the rapid cWB
analysis~\cite{Abbott2016GW150914Minimal}, and GW190521, for which cWB
established a confident detection with minimal waveform
assumptions~\cite{Abbott2020GW190521}. Beyond finding events, waveform
reconstructions from cWB and BayesWave~\cite{Cornish2021BayesWave} are used
in catalog analyses to compare the observed signals with predictions from
binary-black-hole models and investigate possible
discrepancies~\cite{GWTC3}. These contributions demonstrate the value of
unmodeled methods both for detecting astrophysical signals and for
examining their structure without imposing the detailed waveform model
used to interpret the source.

For LISA, identifying an astrophysical transient also requires understanding
how instrumental disturbances enter the data. LISA measures changes in the
phase of laser light exchanged between spacecraft. The raw measurements
are dominated by fluctuations in the laser frequencies. Time-delay
interferometry (TDI) suppresses this noise by combining measurements with
appropriate delays that account for the light travel times between
spacecraft~\cite{Tinto2005TDI,Dhurandhar2002AlgebraicTDI,Vallisneri2005GeometricTDI}.
Other instrumental disturbances are not generally removed by this
cancellation. Transient disturbances, commonly called glitches, can
therefore remain in the TDI outputs and resemble astrophysical signals.
Gravitational waves and glitches can produce different patterns across
these outputs~\cite{RobsonCornish2019Glitches}. In particular, some TDI
combinations have a reduced gravitational-wave response at low frequencies
and can help monitor instrumental noise~\cite{HartwigMuratore2022TDI}.
Although the channels are formed from the same underlying measurements
rather than independent detectors, their relative responses can help
distinguish gravitational waves from glitches. Some disturbances affecting
several laser links can, however, produce similar relative responses in
the channels being compared. In these cases, the channel-response
comparison alone cannot determine whether the candidate is astrophysical
or instrumental.

An unmodeled search could also assist the joint analysis of LISA's
overlapping source populations, usually called the global fit. Existing
pipelines, including GLASS~\cite{LittenbergCornish2023GLASS},
Erebor~\cite{Katz2025Erebor}, and the modular pipeline of Deng et
al.~\cite{Deng2025ModularGlobalFit}, iteratively update their estimates of
massive-black-hole binaries (MBHBs), Galactic binaries, and noise. An
initial search for MBHB signals already forms part of the analysis in
Erebor and the modular
pipeline~\cite{Katz2025Erebor,Deng2025ModularGlobalFit}. An unmodeled search
could support this step by flagging transient candidates for MBHB follow-up
or glitch studies before the search for Galactic binaries, without
requiring them to match a particular source waveform. Their time-frequency
structure, reconstructed waveforms, and relative channel responses could
guide these follow-up studies. Reconstructions could also provide initial
estimates of the transient contributions for subtraction, reducing their
influence on the subsequent Galactic-binary search. These estimates would
then be refined together with the Galactic-binary signals and noise model
during the global fit. Repeating the unmodeled search on the residual data
could reveal transient features missed by the source models.

The fast Wilson--Daubechies--Meyer (WDM) transform described by Necula et
al.~\cite{Necula2012FastWDM} provides the time-frequency representation
used in cWB. For LISA, this representation has supported fast waveform
and likelihood calculations~\cite{Cornish2020TimeFrequency}, studies of
the Galactic foreground~\cite{DigmanCornish2022GalacticBackground}, and
parameter estimation for stellar-origin black-hole
binaries~\cite{DigmanCornish2023StellarBH}. More recent work develops fast
templates for several LISA source classes, including precessing
MBHBs~\cite{Cornish2026FastLISATemplates}. Further developments address
the transform's statistical properties~\cite{Johnson2026WDMFormalism},
differentiable implementations~\cite{Vajpeyi2026DifferentiableWDM},
alternative analysis windows~\cite{Pearson2026CompactTimeFrequency}, and
inference in the presence of data gaps~\cite{PearsonCornish2026DataGaps}.

Searches using time-frequency information include early clustering methods
for extreme-mass-ratio inspirals~\cite{GairJones2007HACRPreprint} and track
finding as the first stage of an MBHB pipeline~\cite{Brown2007ThreeStage}.
Recent examples include searches for bursts from eccentric
binaries~\cite{Knee2024LISABursts} and an excess-power search for premerger
MBHBs, tested on SangriaHM~\cite{Nobili2026FastPremerger}. Semi-coherent
searches also target stellar-mass binary
black holes~\cite{Bandopadhyay2026GlobalTF} and individual harmonics of
extreme-mass-ratio inspirals~\cite{Speri2026SingleHarmonic}. Short Fourier
transforms can accelerate template comparisons, with a pilot application
to MBHB early warning~\cite{TenorioGerosa2025SFT}. Other template-based
methods address rapid MBHB identification~\cite{Cornish2022LowLatency}
and reconstruction for global-fit initialization and early-warning
studies~\cite{Deng2025FastReconstruction}.

Here we present \pversion{}, a pipeline for finding and reconstructing
transient gravitational-wave signals in LISA data without requiring
detailed source waveform templates. The search identifies concentrations
of power in time and frequency and ranks candidates using their strength,
their relative power in two gravitational-wave-sensitive TDI channels,
and their frequency evolution. Selected candidates can be reconstructed,
while comparisons across other TDI combinations help assess a possible
instrumental origin. We test this approach with simulated aligned-spin
and precessing MBHBs, burst signals with different shapes, and instrumental
glitches. We also analyze the Sangria simulated data, recovering all six
injected MBHBs as the six highest-ranked candidates and reconstructing
the signal around each event. The Sangria study demonstrates the search
and reconstruction steps; the response-based glitch studies are performed
separately on controlled simulations. Together, these examples show how
transient finding, response comparisons, and waveform reconstruction can
be combined in a single pipeline.

\section{Setup}
\label{sec:setup}

We first describe the simulation chain, simulator conventions, link-level
responses, and TDI combinations used in the time-frequency analysis.

\subsection{Pipeline overview}
\label{subsec:pipeline-overview}

\begin{figure*}[!t]
\centering
\includegraphics[width=0.84\textwidth]{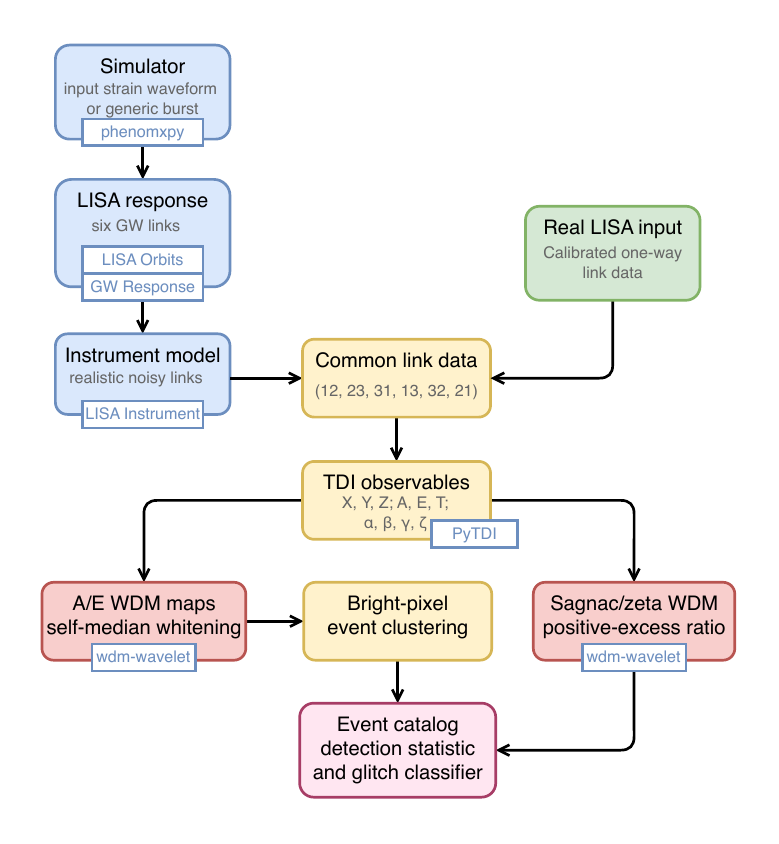}
\caption{Schematic overview of the \pversion{} analysis chain. The simulator branch and the real-data branch both enter through one-way link data before the TDI stage. WDM event clustering, event ranking, and zeta-WDM glitch discrimination are downstream time-frequency data products.}
\label{fig:pipeline-overview}
\end{figure*}

\pversion{} is organized as a sequence of modular functions, shown
schematically in Fig.~\ref{fig:pipeline-overview}. In simulation mode, the user provides a source waveform or a prescribed burst strain time series. The response stage evaluates the six one-way LISA GW links on a common orbit time grid, and the instrument stage adds instrumental noise to the links. The TDI stage constructs the second-generation Michelson, \AET{}, Sagnac, and zeta observables. The search computes WDM maps of the \Achan{} and \Echan{} channels, normalizes them with empirical per-frequency backgrounds, groups connected bright pixels into events, and ranks them. The zeta-WDM stage evaluates the ratio of zeta to excess Sagnac on the same event pixels. For inputs supplied as one-way link data, the downstream analysis begins at the TDI stage. 

\subsection{Simulator}
\label{subsec:simulator}

The simulator starts by generating GW source time series and propagating them
through the LISA response into the six one-way links.  For
massive-black-hole-binary (MBHB) sources, the
strain waveforms are generated with \texttt{phenomxpy}
~\cite{PhenomXPY,GarciaQuirosTiwariBabak2025PhenomXPY}, using the
IMRPhenomTHM model for the aligned-spin
injections~\cite{Estelles2022IMRPhenomTHM}.  The additional precessing examples
use IMRPhenomTPHM as specified in Sec.~\ref{subsec:precessing-validation};
generic response
tests may instead use prescribed burst strain time series.  In both cases the
signal is injected before the TDI combinations are formed.  The generic tests
include sine-Gaussian bursts and white-noise bursts (WNBs).  For the latter,
independent Gaussian-noise series for the two GW polarizations are band-pass
filtered and multiplied by a Gaussian envelope before propagation through the
LISA response.  These are prescribed signal morphologies, distinct from the
instrumental noise added to the link measurements.

All simulations use the time convention of the orbital ephemeris.
The response, instrument, TDI, and WDM calculations share the same time
coordinate, $t_{\rm LISA}$.  The six one-way
links are labelled $(12,23,31,13,32,21)$, where the index pair follows the link
convention of the response and TDI tools.  The response uses LISA Orbits and
LISA GW Response, and the noisy-link simulation uses LISA Instrument
~\cite{BayleHartwig2023Instrument,LISAOrbits,LISAGWResponse,LISAInstrument}.

The instrumental noise is generated by first evolving the LISA Instrument
simulation at $\Delta t=0.1\,{\rm s}$ and evaluating the TDI observables at the
same cadence.  We then remove nonphysical samples: an initial burn-in interval,
which discards startup transients of the simulation, and guard intervals at
chunk boundaries to reduce edge contamination.   After
these samples are removed, the TDI time series are band-passed to
$10^{-5}\le f\le 10^{-1}\,{\rm Hz}$ and downsampled to
$\Delta t=4\,{\rm s}$ for the WDM analysis.  Longer data sets are generated in
overlapped chunks and stitched only after this removal, filtering, and
downsampling procedure has been applied. Standard LISA sensitivity and SNR
conventions are summarized by Babak et al.~\cite{Babak2021Sensitivity}; the
reference-SNR calculation used here is specified below.
The strength and simulated-waveform reconstruction studies use the
fourteen-day noise product. The response comparisons in
Secs.~\ref{subsec:mbhb-glitch-validation}--\ref{subsec:multilink-glitches}
instead use a common, continuously conditioned two-day noise record, whose
padding and analysis interval are specified below. Sangria supplies its own
TDI data and is not generated by this simulation chain.

\subsection{TDI observables}
\label{subsec:tdi-observables}

The TDI evaluations use PyTDI~\cite{PyTDI}.  For the simulated instrument
data, the link measurements are already optical-frequency
fluctuations in Hz, and PyTDI is called with \texttt{unit="frequency"}.
The source-only response path used for the generic and precessing injections
instead forms TDI from fractional-frequency link responses and then multiplies
the resulting time series by the constant laser carrier frequency to express
them in Hz before adding them to the instrument noise products.  This is a
fractional-frequency to frequency conversion after the GW response, not a
direct rescaling of strain.  Sangria retains its supplied dimensionless
fractional-frequency TDI convention.  Throughout this paper, $d_c$ denotes the
TDI time series in channel $c$ in the units of the corresponding study:
Hz for the simulated data and dimensionless fractional frequency
for Sangria. The WDM search performs no conversion between these units.

The default search channels are built from second-generation Michelson
observables $(X_2,Y_2,Z_2)$.  We use the following linear combinations, which
form the standard noise-orthogonal
basis under equal-arm, symmetric-noise assumptions used in LISA sensitivity
studies
~\cite{Prince2002OptimalSensitivity,SGSConventions2026},

\begin{align}
A_2 &= \frac{Z_2-X_2}{\sqrt{2}},\\
E_2 &= \frac{X_2-2Y_2+Z_2}{\sqrt{6}},\\
T_2 &= \frac{X_2+Y_2+Z_2}{\sqrt{3}}.
\label{eq:aet}
\end{align}

For the simulated studies, we use A, E, and T as shorthand for
$A_2$, $E_2$, and $T_2$. For Sangria, the same linear combinations are
applied to its supplied X, Y, and Z observables.

Noise orthogonality is not assumed to hold exactly for the time-dependent
constellation and all noise realizations considered here.  The \Tchan{} channel
is retained as an auxiliary consistency monitor; its GW response is suppressed
in the low-frequency, equal-arm limit, but it is not an exact signal null.  The
\pversion{} event statistic uses self-normalization of only \Achan{} and \Echan{}.

The \pversion{} configuration also evaluates the second-generation Sagnac
channels
$(\alpha_2,\beta_2,\gamma_2)$ and the three second-generation zeta combinations
$(\zeta_{21},\zeta_{22},\zeta_{23})$ available in the TDI package used. These combinations are part of the algebraic
and second-generation TDI framework, including Sagnac/zeta noise-monitoring
constructions
~\cite{Dhurandhar2002AlgebraicTDI,Muratore2022NoiseMonitors,
HartwigMuratore2022TDI,TintoDhurandharMalakar2023G2TDI}.

\section{Time-Frequency Representation}
\label{sec:wdm}

\subsection{WDM transform}
\label{subsec:wdm-transform}

For a TDI time series $d_c[k]$ in channel $c$, the WDM coefficient map is
computed as
\begin{equation}
W_c(t_i,f_j)
=
\mathcal{W}_{M,K,\beta,p}\!\left[d_c\right](t_i,f_j),
\qquad c\in\mathcal{S},
\label{eq:wdm-coeff}
\end{equation}
where $\mathcal{W}_{M,K,\beta,p}$ denotes the Wilson--Daubechies--Meyer
transform
implemented by the \texttt{wdm-wavelet} package \cite{WDMWaveletPackage}, which is part of the \texttt{pycwb} infrastructure \cite{pycwb}, and $\mathcal{S}$ denotes the
TDI channels transformed in a given analysis.  Standard cWB analyses combine
multiple WDM resolutions to better match signals with different time-frequency
morphologies.  The reference \pversion{} search uses a single resolution,
$M=512$, $K=32$, \texttt{beta\_order}=6, and \texttt{precision}=6; combining multiple resolutions is left for future
investigation.  Finding a candidate and recovering its waveform are kept as separate tasks.
The search locates a promising region in time and frequency; reconstruction
then returns to the observed data to recover the signal in that region.
The transform is evaluated separately for these two tasks.
The reference search uses $K=32$, while the reconstruction examples use
$K=256$.  These are empirical working configurations for the examples studied
here, not the outcome of a systematic optimization of $K$ or uniquely
preferred values.  Further optimization can be investigated for particular
signals and analysis requirements.  The separate simulated reconstruction
study in Sec.~\ref{subsec:simulated-reconstruction} has its own selection
settings; the targeted follow-ups start from the discovered masks and grow
connected energetic regions for reconstruction.  At the downsampled cadence
$\Delta t=4\,{\rm s}$ this gives a frequency spacing
$\Delta f=2.44140625\times10^{-4}\,{\rm Hz}$ in the coefficient maps and a
median WDM
time-bin spacing of approximately $2.05\times10^3\,{\rm s}$.  The analysis band
starts at $10^{-5}\,{\rm Hz}$, but the WDM maps in
Figs.~\ref{fig:primary-validation-summary} and \ref{fig:generic-burst}
start at the first nonzero
frequency bin available in the map, $2.44140625\times10^{-4}\,{\rm Hz}$, and
extend to $2\times10^{-2}\,{\rm Hz}$.  This time-frequency representation
follows the fast WDM transform described in
Refs.~\cite{Necula2012FastWDM,Klimenko2016TransientMethod}; the Python
implementation used here is \texttt{wdm-wavelet}~\cite{WDMWaveletPackage}.

A time-frequency pixel denotes one indexed pair $(t_i,f_j)$ on the map.  It
carries the complex coefficient $W_c(t_i,f_j)$; the corresponding raw power is
$|W_c(t_i,f_j)|^2$.

\subsection{Whitening}
\label{subsec:whitening}

We divide channel power by a frequency-dependent empirical noise floor to
compare excesses within the event map.  The default
\pversion{} whitening is the self-A/E median normalization, using the analyzed
data segment itself to estimate this background.  The
background is estimated separately in each frequency bin, using the median over
time rather than the mean.  This makes the normalization less sensitive to a
small number of very loud transient pixels while still tracking the typical
frequency-dependent channel power.  For $c\in\{A_2,E_2\}$, define the raw WDM
power and frequency-wise empirical noise floor as

\begin{align}
Q_c(t_i,f_j) &= |W_c(t_i,f_j)|^2,\\
B_c(f_j) &= {\rm median}_{i}\, Q_c(t_i,f_j).
\label{eq:self-ae-median}
\end{align}

The whitened coefficients and powers are
\begin{align}
\widehat{W}_c(t_i,f_j) &= \frac{W_c(t_i,f_j)}{\sqrt{B_c(f_j)}},\\
P_c(t_i,f_j) &= |\widehat{W}_c(t_i,f_j)|^2
             = \frac{Q_c(t_i,f_j)}{B_c(f_j)}.
\label{eq:self-ae-power}
\end{align}
The event-clustering stage uses $P_{\rm AE}=P_{A_2}+P_{E_2}$.
It should be noted that this rescaling does not establish independent coefficients or a unit-variance Gaussian background by itself.

\section{Event Clustering and Detection Statistic}
\label{sec:event-statistic}

\subsection{Bright-pixel event clustering}
\label{subsec:event-finder}

The clustering stage groups bright pixels in the whitened WDM map into
candidate events.  It operates on the combined search power
$P_{\rm AE}$ in a chosen time window and frequency band.  The loudest pixels,
defined as those at or above the 99th percentile of $P_{\rm AE}$ in the chosen search window and band, are used as \textit{seeds}. Pixels at or above the lower 95th percentile threshold are allowed to join a seed only if they are locally connected to it. The two thresholds allow a high-power seed to retain connected, lower-power
pixels in the same time-frequency structure.

Seeds are processed in order of decreasing power.  Each component grows
according to a local connectivity rule.  Within one WDM time bin, each
connection joins adjacent frequency bins.  Between neighboring WDM time bins,
each connection permits a separation of at most four frequency bins.  These
are per-connection limits: repeated connections can produce a component with
a larger total frequency extent.  They permit chirp-like tracks to form a
single event while excluding pixels without a connecting path above the
growth threshold.  Components with fewer than eight pixels or fewer than two
occupied time bins are discarded.  Each surviving connected component defines an
event pixel set $\mathcal{C}$ that is passed to the event statistic and to the
zeta-WDM glitch classifier. The choices made here for the sparseness and the connectivity of the pixels are based on empirical tests done for a small set of MBHB injections, and can be improved and targeted to other signal classes. 

\subsection{Pipeline detection statistic}
\label{subsec:d-stat}

The ranking combines the total power in a cluster, its balance between
\Achan{} and \Echan{}, and the evolution of its centroid frequency.  For an
event pixel set $\mathcal{C}$, the channel energies are
\begin{equation}
E_A=\sum_{(i,j)\in\mathcal{C}}P_{A_2}(t_i,f_j),
\qquad
E_E=\sum_{(i,j)\in\mathcal{C}}P_{E_2}(t_i,f_j),
\label{eq:event-channel-energies}
\end{equation}
with

\begin{align}
E_{\rm AE} &= E_A + E_E,\\
B_{\rm AE} &= \frac{4E_AE_E}{(E_A+E_E)^2}.
\label{eq:event-balance}
\end{align}

The balance factor is unity when the event energy is equally split between
\Achan{} and \Echan{} and tends to zero when one channel dominates.

To quantify frequency evolution, we first compute the energy-weighted centroid
frequency in each occupied time bin,
\begin{equation}
\bar f(t_i)
=
\frac{\sum_j f_j\,P_{\rm AE}(t_i,f_j)\,
{\bf 1}_{(i,j)\in\mathcal{C}}}
{\sum_j P_{\rm AE}(t_i,f_j)\,{\bf 1}_{(i,j)\in\mathcal{C}}},
\label{eq:centroid-frequency}
\end{equation}
and
\begin{equation}
m_{\rm chirp}
=
\frac{1}{N_t-1}
\sum_{\ell=1}^{N_t-1}
{\bf 1}\!\left[
\bar f(t_{\ell+1})-\bar f(t_\ell)
\ge -\Delta f
\right],
\label{eq:chirp-monotonicity}
\end{equation}
where $N_t$ is the number of occupied WDM time bins, indexed in time order.
The indicator $\mathbf{1}$ is one when the stated condition holds and zero
otherwise.  Thus $m_{\rm chirp}$ is the fraction of successive occupied bins
whose centroid frequency does not decrease by more than one WDM frequency bin.
The ranking statistic is
\begin{equation}
\rho_{\rm event}
=
\sqrt{E_{\rm AE}}\,\sqrt{B_{\rm AE}}\,
\left(\frac{1+m_{\rm chirp}}{2}\right).
\label{eq:rho-event}
\end{equation}
The three factors combine normalized cluster power, channel balance, and a
preference for approximately nondecreasing centroid frequency. The chirp term is a preference in the ranking, not a strict requirement on
candidate morphology: its multiplicative factor lies between $1/2$ and $1$,
so a nonchirping event is not rejected solely for lacking an increasing
frequency.  Other target morphologies can be accommodated by replacing or
omitting this weighting in an alternative ranking statistic.  Such a change
would affect candidate ordering and require its own performance assessment;
all results here use Eq.~(\ref{eq:rho-event}), including the generic-burst
examples presented below. 

For comparisons against same-search noise fluctuations, we also define an
empirical noise scale
\begin{equation}
\rho_{\rm noise}
=
{\rm median}_{k}
\left[
\max_{\mathcal{C}\in\mathcal{W}_k}
\rho_{\rm event}(\mathcal{C})
\right],
\label{eq:rho-noise}
\end{equation}
where $\mathcal{W}_k$ are off-source noise windows with the same duration,
frequency band, WDM resolution, and event-clustering settings as the signal
search window.  The corresponding rho-ratio is
\begin{equation}
R_\rho = \frac{\rho_{\rm event}}{\rho_{\rm noise}}.
\label{eq:rho-ratio}
\end{equation}
This ratio is an empirical normalization for comparing event strengths in the
validation study; it is not a calibrated false-alarm probability.

\begin{figure}[t]
\centering
\includegraphics[width=0.48\textwidth]{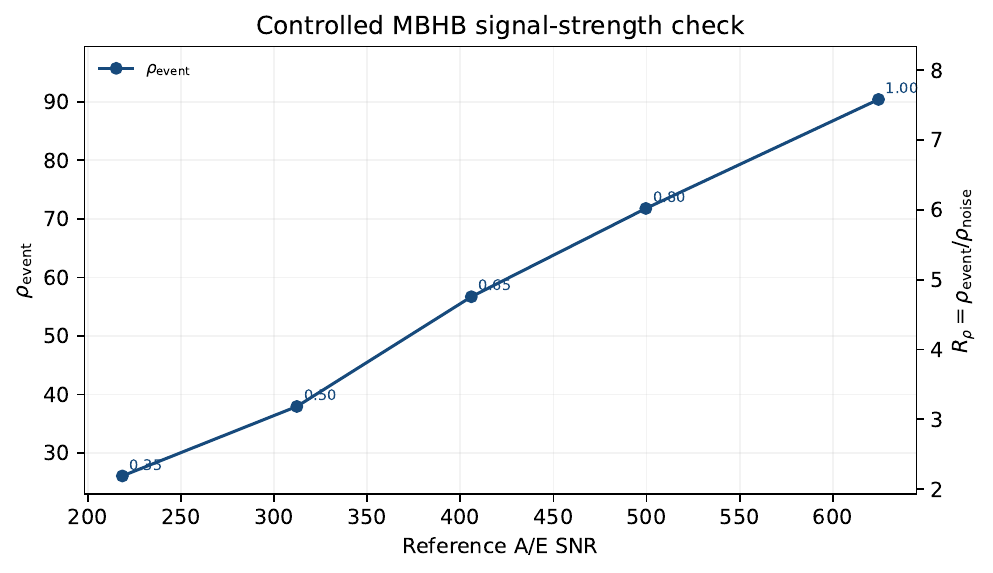}
\caption{
Reference A/E SNR (under the
diagonal-channel noise approximation described in the text) versus recovered $\rho_{\rm event}$ and
$R_\rho=\rho_{\rm event}/\rho_{\rm noise}$ for five distance-scaled versions of
the same MBHB response.  The left axis gives $\rho_{\rm event}$ and the right
axis gives $R_\rho$.  Point labels give the amplitude scale relative to the
reference $D_L=6230\,{\rm Mpc}$ source.
}
\label{fig:snr-rho-event}
\end{figure}

To check that Eq.~(\ref{eq:rho-event}) responds monotonically to signal
strength,
we rescale the stored response of a $3\times10^5\,M_\odot$, $q=0.6$ MBHB before
adding it to the same realistic-noise \AET{} streams.  This is equivalent to
changing the luminosity distance while holding the source orientation, sky
position, and phase evolution fixed.  For each scale, the reference SNR is
computed from the sum of the \Achan{} and \Echan{} TDI inner products over
$10^{-5}\le f\le10^{-1}\,{\rm Hz}$. Each channel uses a one-sided noise
power spectral density estimated with median Welch averaging and Hann
windows. This is the optimal SNR under the diagonal-channel noise
approximation; cross-spectral covariance between A and E is neglected.
The recovered $\rho_{\rm event}$ and
$R_\rho$ increase monotonically with this reference SNR over the tested range, as
shown in Fig.~\ref{fig:snr-rho-event}.

For the same event pixels, we also compute the complex coherence between
the \Achan{} and \Echan{} WDM coefficients,
\begin{equation}
C_{\rm AE}
=
\frac{\left|\sum_{(i,j)\in\mathcal{C}}
\widehat W_{A_2}(t_i,f_j)\widehat W^*_{E_2}(t_i,f_j)\right|}
{\sqrt{E_AE_E}}.
\label{eq:ae-coherence}
\end{equation}
This quantity measures the complex channel consistency of the event and is
reported in the catalog but does not enter $\rho_{\rm event}$.

\begin{figure}[t]
\centering
\includegraphics[width=0.48\textwidth]{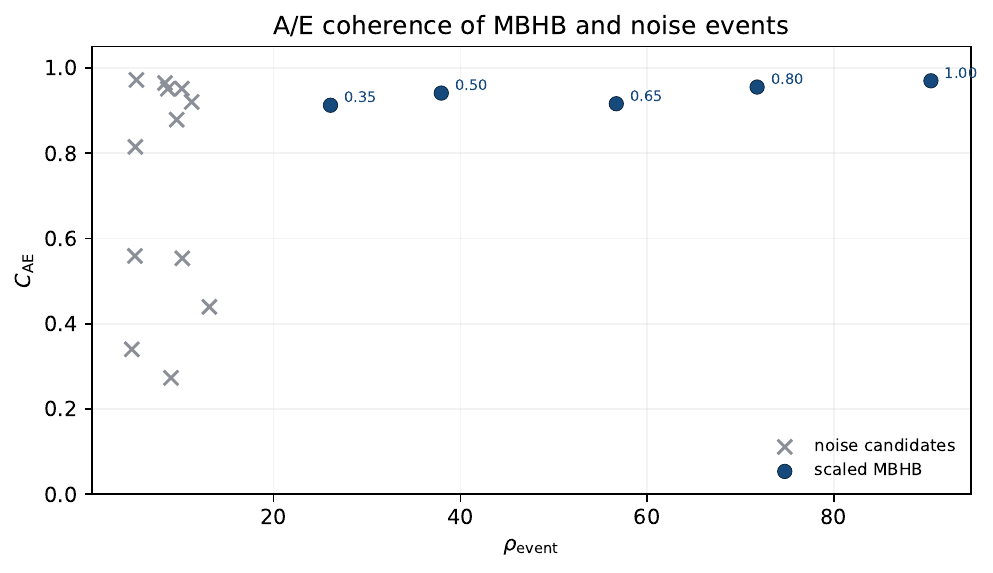}
\caption{
A/E complex coherence for distance-scaled MBHB events and randomly sampled
off-source noise candidates selected with the same event finder.  The horizontal
axis is the event ranking statistic $\rho_{\rm event}$ and the vertical axis is
$C_{\rm AE}$.  In this controlled sample, the MBHB points remain above
$C_{\rm AE}\simeq0.8$, while the noise candidates show a broader scatter.  Point
labels on the MBHB events give the amplitude scale relative to the reference
$D_L=6230\,{\rm Mpc}$ source.
}
\label{fig:ae-coherence-mbhb-noise}
\end{figure}

Figure~\ref{fig:ae-coherence-mbhb-noise} shows the same scaled MBHB events and a
deterministic random sample of off-source noise candidates in the
$(\rho_{\rm event},C_{\rm AE})$ plane.  The MBHB events in this controlled test
remain highly coherent between \Achan{} and \Echan{}, while the sampled noise
events span a broader range of coherence.  In this
sample the MBHB events have $C_{\rm AE}\gtrsim0.8$, whereas the noise candidates
span from weak to strong apparent coherence.

\section{Zeta-WDM Glitch Classifier}
\label{sec:zeta-wdm}

The \Achan{}/\Echan{} statistic above selects time-frequency excess using
power, channel balance, and frequency evolution.  It is not by itself a glitch
classifier as it can assign high values to instrument glitches.  Local
instrumental transients can produce bright WDM structures that pass an
excess-power trigger, especially when the transient is localized to one optical
bench, test mass, or inter-spacecraft link.  Such effects are expected to be an
important part of LISA transient searches
~\cite{RobsonCornish2019Glitches,Baghi2022LPFGlitches,
Spadaro2023GlitchSystematics,Muratore2025GlitchPipeline}.  We therefore add a
second, event-level consistency test using Sagnac and zeta TDI combinations,
which have been studied as instrument-noise monitors for LISA
~\cite{Muratore2022NoiseMonitors,HartwigMuratore2022TDI} and are evaluated here
with the same second-generation TDI framework used for the Michelson channels
~\cite{Tinto2005TDI,TintoDhurandharMalakar2023G2TDI}.

The zeta-WDM study compares the power on the recovered event pixels in
Sagnac-like and zeta monitor channels. Its background estimate is obtained
from the observed data. For each channel,
\begin{equation}
c\in\{\alpha_2,\beta_2,\gamma_2,\zeta_{21},\zeta_{22},\zeta_{23}\},
\end{equation}
we define $Q_c^{\rm obs}(t_i,f_j)=|W_c^{\rm obs}(t_i,f_j)|^2$.
We denote by $\mathcal U$ the union of the pixel masks of all candidates
retained by the A/E search. The frequency-dependent background is
\begin{equation}
B_c(f_j)=\mathop{\rm median}_{i:(i,j)\notin\mathcal U}
Q_c^{\rm obs}(t_i,f_j).
\label{eq:zeta-floor}
\end{equation}
Thus every channel uses the same exclusions, set by the A/E search,
while retaining its own frequency-dependent normalization. We define
\begin{equation}
X_c(t_i,f_j)=\max\left[\frac{Q_c^{\rm obs}(t_i,f_j)}{B_c(f_j)}-1,\,0\right].
\label{eq:positive-excess}
\end{equation}
This quantity measures positive power excess above a median background.
It does not subtract the noise fluctuation realized in an individual pixel:
positive noise fluctuations remain after clipping. Excluding candidate pixels
reduces direct event contamination of the estimate. The median uses the
remaining time bins across the observed record, rather than a local window
around each event. A channel-frequency
estimate requires a finite, positive median from the remaining samples;
an event ratio with a zero denominator is undefined. Summing over the
previously found event pixels gives
\begin{align}
E_{\alpha\beta\gamma}(\mathcal{C})
&=
\sum_{c\in\{\alpha_2,\beta_2,\gamma_2\}}
\sum_{(i,j)\in\mathcal{C}}X_c(t_i,f_j),\\
E_{\zeta}(\mathcal{C})
&=
\sum_{c\in\{\zeta_{21},\zeta_{22},\zeta_{23}\}}
\sum_{(i,j)\in\mathcal{C}}X_c(t_i,f_j),
\label{eq:zeta-excess-sums}
\end{align}
and the event-level zeta-WDM ratio is

\begin{equation}
R_{\zeta}
=
\frac{E_{\zeta}(\mathcal{C})}{E_{\alpha\beta\gamma}(\mathcal{C})}.
\label{eq:zeta-ratio}
\end{equation}

The calculation proceeds from the channel-wise excess $X_c$ to the summed
Sagnac and zeta excesses and finally to $R_\zeta$. A small ratio means less
selected, normalized excess in the zeta channels relative to the Sagnac
channels. We use $R_\zeta$ as a response-consistency statistic for glitch
studies, not as a calibrated veto; it does not enter the event ranking used
here. Sections~\ref{subsec:mbhb-glitch-validation}
and \ref{subsec:generic-burst-validation} examine its behavior for prescribed
GW and single-link injections and Sec.~\ref{subsec:multilink-glitches} tests
more complex readout patterns.

\section{Validation Study}
\label{sec:validation}
The validation addresses three connected questions: which candidates are
found, what their channel responses reveal, and what waveform information can
be recovered. Table~\ref{tab:validation-guide} maps these questions to the
individual studies and selection procedures. The MBHB and generic-burst
studies examine candidate finding and TDI response comparisons for prescribed
GW and link-glitch injections. The Sangria analysis instead checks the
leading candidates against the revealed MBHB injections. It uses the supplied
\Achan{}/\Echan{} data and does not evaluate the zeta response comparison,
because the required link data are unavailable in that product.
Reconstruction studies assess waveform extraction separately, with selection
settings distinguished from candidate identification.

OpenAI Codex (GPT-6 Astra) assisted in the development and checking of scripts for some of the validation, benchmarking, and plotting studies under the authors' direction.

\begin{table*}[t]
\caption{Guide to the validation studies.  Candidate finding, response
interpretation, and waveform recovery answer different questions; the
selection and truth information used in each study are therefore stated
separately.  Detailed numerical comparisons appear at the cited locations.}
\label{tab:validation-guide}
\small
\begin{ruledtabular}
\begin{tabular}{p{0.16\textwidth}p{0.27\textwidth}
  p{0.24\textwidth}p{0.23\textwidth}}
Study & What is tested? & Where to find it & What the test does not establish \\
\hline
Black-hole signals and disturbances & Can the same search recover the
injected source and characterize its response? Uses recovered candidates and
an observed-data background estimate. & Sec.~\ref{subsec:mbhb-glitch-validation};
Fig.~\ref{fig:primary-validation-summary}. & A shared instrumental-noise
realization and prescribed sources; no population recovery measurement. \\
Other burst signals & Does the procedure recover other signal shapes?
Same search and response statistic for every record. &
Sec.~\ref{subsec:generic-burst-validation}; Fig.~\ref{fig:generic-burst};
Table~\ref{tab:signal-recovery}. & A finite set of relocated burst examples;
no test of unresolved foregrounds or data gaps. \\
Instrument disturbances & How do different link patterns and strengths affect
the response statistic? Reference and loud-glitch trials use the leading
candidate. & Sec.~\ref{subsec:multilink-glitches};
Fig.~\ref{fig:multilink-current}. & Shared noise does not provide independent
background trials; no calibrated rejection rate or veto threshold. \\
Waveform recovery & How much of a known signal can be reconstructed? Search
intervals use the known injection times. &
Sec.~\ref{subsec:simulated-reconstruction};
Fig.~\ref{fig:observed-reconstruction}. & Selection uses injection information;
this tests waveform extraction rather than blind candidate finding. \\
Precessing black holes & Can the search find and reconstruct more complex
signals? Reconstruction follows the leading search candidate. &
Sec.~\ref{subsec:precessing-validation};
Fig.~\ref{fig:precessing-reconstruction};
Table~\ref{tab:precessing-reconstruction}. & Only three development examples;
expanding the recovered region improves two and worsens one. \\
Sangria simulation & What does the search find in a year of challenge data?
Leading candidates are checked against the revealed injected signals and then
reconstructed. & Sec.~\ref{subsec:sangria-results}; Figs.~\ref{fig:sangria-demo}
and \ref{fig:sangria-targeted-combined}. & No instrument-monitor comparison;
reference combines all black-hole signals; local recovery is not an
isolated-source full-waveform test. \\
\end{tabular}
\end{ruledtabular}
\end{table*}

\subsection{MBHB-plus-glitch validation}
\label{subsec:mbhb-glitch-validation}

This study uses a continuously generated two-day instrumental-noise
record, sampled initially at $0.1\,\mathrm{s}$. All Michelson and monitor
channels are derived from the same instrument realization and the same ESA
trailing orbit, sampled every $600\,\mathrm{s}$. After an initial
$600\,\mathrm{s}$ is discarded due to edge effects, the TDI series retains $3000\,\mathrm{s}$ of
padding at each end during the $10^{-5}$--$10^{-1}\,\mathrm{Hz}$ bandpass
and downsampling to $4\,\mathrm{s}$. The record is cropped only
after this conditioning.

Source responses are generated and conditioned over longer seven- or fourteen-day records before the common two-day interval is extracted; the continuous padded conditioning described above applies to the instrumental noise. Each source is added separately to this common instrumental-noise record. These response tests do not include a Galactic-binary foreground, data gaps, or several simultaneous injected sources. GW reference peaks are placed $0.5$ day after the record start, and glitch reference onsets $1.5$ days after the record start; relative component delays are preserved.
The search covers $0.25$--$1.75$ days after the record start, with $M=512$, $K=32$, the $10^{-5}$--$0.02\,\mathrm{Hz}$ band, seed/grow quantiles $0.99/0.95$, and minimum sizes of eight pixels and two time bins as stated before. Up to 30 candidates are retained. Injection times do not select event pixels or rescue candidates below these cuts. Background medians in Eq.~(\ref{eq:zeta-floor}) use the two-day observed data outside the union
of all retained candidate masks. The same prescription is used for every source and amplitude level. The injection parameters are collected in Table~\ref{tab:injection-parameters}. The MBHB polarizations are produced with \texttt{phenomxpy} and propagated through the time-dependent LISA response before TDI. 

The glitches are rectangular \texttt{lisaglitch}~\cite{LISAGlitch} disturbances in the readouts and are listed in Table~\ref{tab:injection-parameters}. The glitch studies are motivated by earlier LISA transient analyses
\cite{RobsonCornish2019Glitches,Baghi2022LPFGlitches,
Spadaro2023GlitchSystematics,Muratore2025GlitchPipeline}.

\begin{table*}[t]
\caption{Injection parameters for the MBHB, generic-burst and single-link
studies. Every injection is added separately to the same two-day noise
data stream. GW peak references are at 0.5 day and glitch onsets at 1.5 days
from its start. The MBHB masses are detector-frame totals, $q=m_2/m_1$, and $s_1,s_2$
are the dimensionless spin components along the orbital angular momentum.
All three use $D_L=6230\,\mathrm{Mpc}$, effective inclination zero and waveform generation
from $10^{-4}\,\mathrm{Hz}$. Sky angles and phases are in radians; the sky
entries are the $(\mathrm{ra},\mathrm{dec})$ inputs to the LISA response.
Burst width means $6\sigma$ of the Gaussian envelope, and strain amplitude
is the peak combined-polarization amplitude.
SG denotes sine-Gaussian and WNB white-noise burst. The final block gives
rectangular readout-glitch levels; the loud levels follow the fixed-$Q$
prescription in the text. Dashes denote inapplicable
parameters.}
\label{tab:injection-parameters}
\begin{ruledtabular}
\begin{tabular}{lrrrrr}
MBHB mass [$M_\odot$] & $q$ & Sky angles & $s_1$ & $s_2$ & $\phi_{\rm ref}$ \\
\hline
$10^5$ & 1 & (0.45, 0.35) & 0.1 & -0.1 & 0.2 \\
$3\times10^5$ & 0.6 & (1.75, -0.45) & -0.2 & 0.2 & 1.1 \\
$10^6$ & 0.3 & (3.1, 0.1) & 0.3 & -0.3 & 2.4 \\
\end{tabular}
\par\smallskip
\begin{tabular}{lrrrrrr}
Burst & Frequency/band [mHz] & Width [s] & Strain [$10^{-18}$] & Sky angles & Phase & Seed \\
\hline
SG 01 & 2.2 & 9000 & 8 & (0.45, 0.35) & 0.3 & \textemdash \\
WNB 01 & 1.2--6 & 11000 & 7 & (1.55, -0.4) & \textemdash & 1101 \\
SG 02 & 7.5 & 6500 & 8 & (3.1, 0.12) & 1.4 & \textemdash \\
WNB 02 & 3--12 & 8000 & 7.5 & (4.6, -0.62) & \textemdash & 1102 \\
\end{tabular}
\par\smallskip
\begin{tabular}{llrrr}
Study & Readout & Duration [s] & Reference level [Hz] & Loud level [Hz] \\
\hline
Primary & 12 & 1400 & $1.8\times10^{-4}$ & $1.295\times10^{-3}$ \\
Primary & 23 & 2600 & $1.2\times10^{-4}$ & $6.297\times10^{-3}$ \\
Primary & 31 & 1000 & $3\times10^{-4}$ & $2.186\times10^{-3}$ \\
Generic & 12 & 5500 & $2\times10^{-4}$ & $1.294\times10^{-3}$ \\
Generic & 31 & 2500 & $3\times10^{-4}$ & $5.124\times10^{-3}$ \\
\end{tabular}
\end{ruledtabular}
\end{table*}

The listed glitch amplitudes define the reference injections. Each glitch
is also tested as a loud injection, with its amplitude fixed by the following
power normalization.
We define its injected A/E power $Q$ by summing the source-only WDM power,
normalized separately by each channel's simulated-noise median, over the
search band and interval. A single factor $\sqrt{Q_*/Q}$ multiplies all
components of a glitch, with $Q_*=1.053584606\times10^6$ fixed by a separate
$10^6\,M_\odot$, $q=0.6$, face-on reference MBHB injection. This gives each
loud glitch the same injected A/E power as that reference. 

\begin{figure*}[t]
\centering
\includegraphics[width=0.96\textwidth]{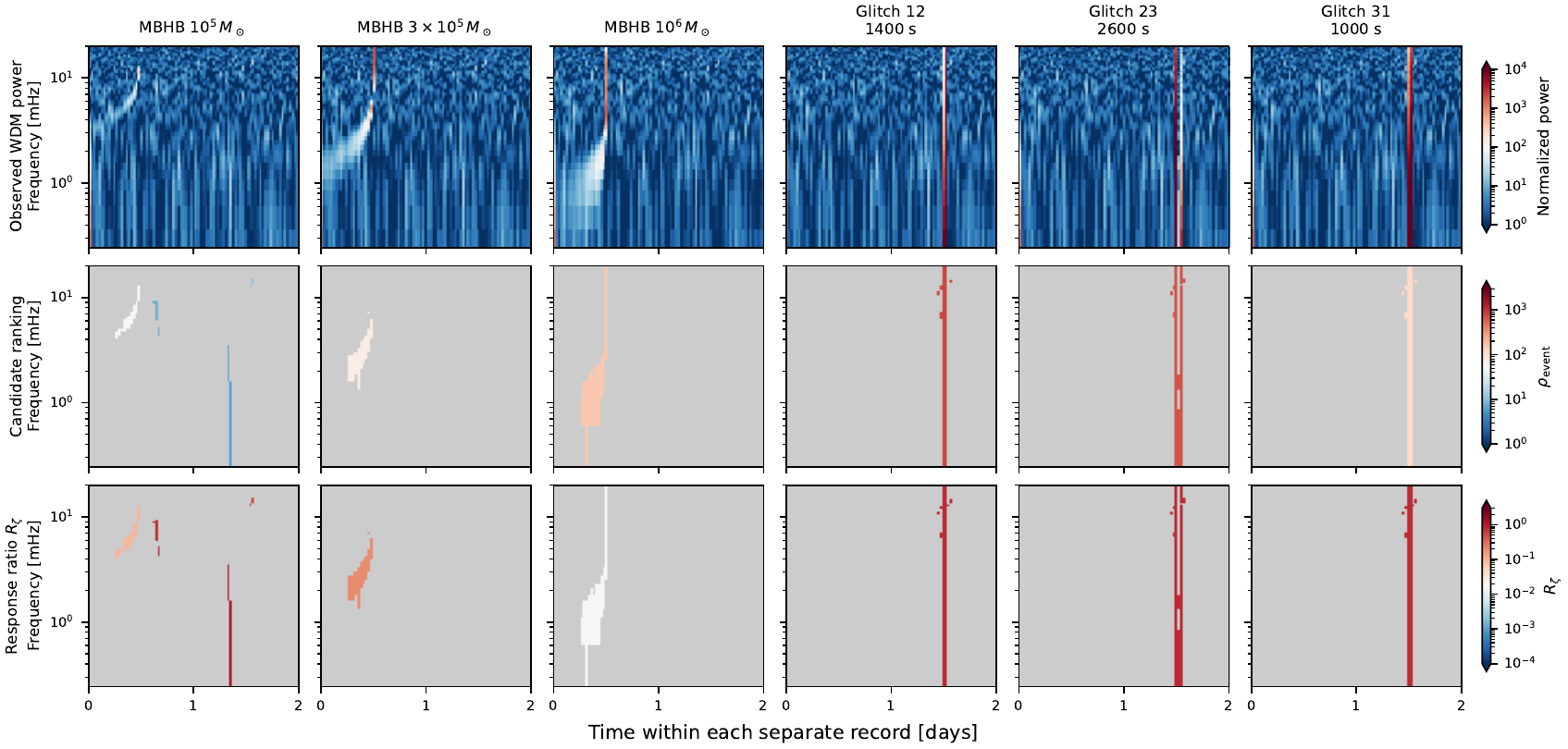}
\caption{ Time-frequency view of the MBHB and single-link glitch study. Each column is a separate two-day source-plus-noise record. GW signals use their reference amplitudes and glitches use the loud-injection levels in Table~\ref{tab:injection-parameters}. Top: observed A+E power, with each channel normalized by its frequency-wise median over the simulated data period. Middle: each retained candidate's ranking statistic is assigned to its selected A/E cluster. Bottom: the candidate's integrated $R_\zeta$ is assigned to those same clusters. The lower rows show all retained candidates, not only
the leader; grey regions contain no selected pixels. These event-level
values are constant over each candidate mask and are not pixel-wise
statistics or a pixel-wise zeta fraction. The full two-day record is shown
from the first nonzero WDM frequency bin to $0.02\,\mathrm{Hz}$;
the search band starts at $10^{-5}\,\mathrm{Hz}$. Frequency and colour scales are logarithmic, with common colour limits across the two figures and endpoint colours for values beyond the limits. Plotting code was developed with assistance from OpenAI Codex (GPT-6 Astra).}
\label{fig:primary-validation-summary}
\end{figure*}

Figure~\ref{fig:primary-validation-summary} shows the outcome of the
search on the MBHB and glitch injections, from the observed
time-frequency power to the selected candidates and their response
ratios. For the first MBHB injection, the search recovers a
signal-bearing track together with additional clusters produced by
noise fluctuations. The MBHB candidate has a higher ranking statistic
($\rho_{\rm event}=54.2$) than these noise clusters
($\rho_{\rm event}=6.5$--$11.1$), and a smaller response ratio:
$R_\zeta=0.097$, compared with $0.475$--$1.024$ for the fluctuations.
The other MBHB examples likewise have smaller response ratios than
the single-link glitches shown here. These examples illustrate the
intended role of the analysis: identify transient power and then use
the detector response to help assess its origin.
Table~\ref{tab:signal-recovery} summarizes the numerical comparisons. 




\subsection{Generic burst-plus-glitch validation}
\label{subsec:generic-burst-validation}

The generic-burst study uses the two sine-Gaussian and two white-noise-burst
injections listed in Table~\ref{tab:injection-parameters}. For the WNBs,
a fifth-order Butterworth bandpass is applied forwards and backwards to the
seeded random series before the Gaussian envelope. The envelope widths and frequency bands do not define strict support boundaries. The combined-polarization amplitude is normalized within three envelope standard deviations.

\begin{figure*}[t]
\centering
\includegraphics[width=0.96\textwidth]{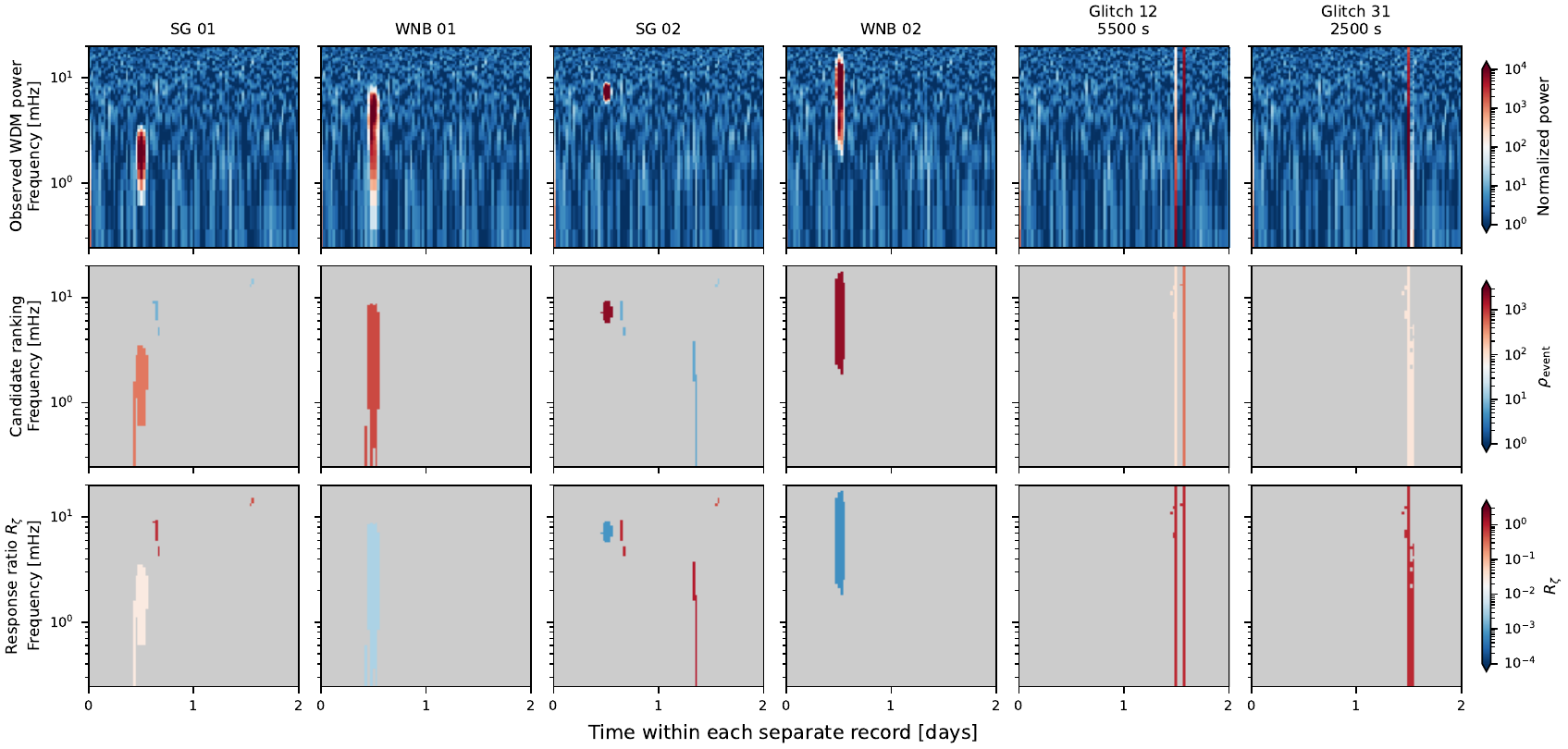}
\caption{Full time-frequency view of the generic-burst and associated glitch
study. Each column is a separate two-day source-plus-noise record, using
the GW reference amplitudes and loud-glitch levels in
Table~\ref{tab:injection-parameters}. The rows show observed power, candidate
ranking and integrated $R_\zeta$, following the definitions and common
colour scales of Fig.~\ref{fig:primary-validation-summary}. All retained
candidates are shown; the lower rows assign each event's scalar statistic
to its selected pixels, with grey elsewhere. Plotting code was developed
with assistance from OpenAI Codex (GPT-6 Astra).}
\label{fig:generic-burst}
\end{figure*}

The two associated glitch configurations are also listed in
Table~\ref{tab:injection-parameters}. Each signal or glitch is searched in
its own simulated source-plus-noise record. Figure~\ref{fig:generic-burst} presents the search outputs for the sine-Gaussian and white-noise-burst injections, together with the associated glitches. The search identifies the injected bursts despite their different time-frequency structures.
Their leading candidates have $R_\zeta$ between $7.25\times10^{-4}$ and $0.0272$, whereas the glitch candidates shown here have ratios of approximately $0.81$--$0.83$. Additional noise clusters in the sine-Gaussian data have lower ranking statistics and larger response ratios than the injected bursts. These examples demonstrate the expected response contrast beyond chirping MBHB signals.

\begin{table*}[t]
\caption{Leading-candidate metrics for the primary and generic studies.
The search statistic and coherence are shown for the reference injections.
The two ratio columns compare reference and loud-glitch injections, with
the loud-injection amplitude defined in the text.
GW signals are tested only at their specified amplitudes. MBHB masses are
in $M_\odot$. A dash denotes no retained
candidate, an undefined ratio, or an inapplicable amplitude level. All quantities use recovered event masks and the same observed-data background prescription.}
\label{tab:signal-recovery}
\begin{ruledtabular}
\begin{tabular}{lcccc}
Case & $\rho_{\rm event}$ & $C_{\rm AE}$ & $R_\zeta$ (reference) & $R_\zeta$ (loud glitches) \\
\hline
\multicolumn{5}{c}{MBHB and primary glitches} \\
MBHB $10^5$, q=1.0 & 54.2 & 0.924 & 0.0967 & \textemdash \\
MBHB $3\times10^5$, q=0.6 & 73.1 & 0.97 & 0.201 & \textemdash \\
MBHB $10^6$, q=0.3 & 163 & 0.966 & 0.0165 & \textemdash \\
Link 12, 1400 s & 120 & 0.945 & 0.835 & 0.815 \\
Link 23, 2600 s & 11.1 & 0.965 & 0.477 & 0.805 \\
Link 31, 1000 s & 29.5 & 0.383 & 0.816 & 0.802 \\
\multicolumn{5}{c}{Generic bursts and glitches} \\
SG 01 & 459 & 0.997 & 0.0272 & \textemdash \\
WNB 01 & 764 & 0.314 & $3.35\times10^{-3}$ & \textemdash \\
SG 02 & $1.97\times10^3$ & 0.998 & $8.34\times10^{-4}$ & \textemdash \\
WNB 02 & $1.83\times10^3$ & 0.179 & $7.25\times10^{-4}$ & \textemdash \\
Link 12, 5500 s & 75 & 0.954 & 0.816 & 0.824 \\
Link 31, 2500 s & 11.1 & 0.964 & 0.475 & 0.829 \\
\end{tabular}
\end{ruledtabular}
\end{table*}

\subsection{Single- and multi-link glitch response}
\label{subsec:multilink-glitches}

The twelve configurations in Table~\ref{tab:multilink-injections} test how
link geometry and amplitude affect the search. They comprise
six single-link disturbances, three pairs, two cyclic triplets and one
simultaneous all-six-link disturbance. Each configuration is tested at both
reference and loud amplitudes, with component timings and relative
amplitudes preserved.

\begin{table*}[t]
\caption{Single- and multi-link glitch injection parameters. All components
are rectangular disturbances in the listed directed readouts. Within a row,
one value applies to every component; tuples follow the listed readout
order. Onset offsets are measured from 1.5 days after the start of each
two-day simulated data stream. The same amplitude factor converts
all components from reference to loud levels using the power normalization
in Sec.~\ref{subsec:mbhb-glitch-validation}. The simultaneous reciprocal pair and all-six-link case are the two configurations separated in the lower panel of Fig.~\ref{fig:multilink-current}.}
\label{tab:multilink-injections}
\begin{ruledtabular}
\begin{tabular}{lrrrr}
Readouts & Width [s] & Onset offsets [s] & Reference level [$10^{-4}$ Hz] & Loud level [Hz] \\
\hline
12 & 600 & 0 & 3.5 & $6.191\times10^{-3}$ \\
23 & 4000 & 0 & 0.8 & $3.308\times10^{-3}$ \\
31 & 1400 & 0 & 1.8 & $1.043\times10^{-3}$ \\
13 & 1000 & 0 & 4 & $2.187\times10^{-3}$ \\
21 & 3000 & 0 & 1 & $3.133\times10^{-3}$ \\
32 & 1800 & 0 & 2.2 & $2.069\times10^{-3}$ \\
12+21 & 1500 & 0 & 1.5 & $6.556\times10^{-4}$ \\
12+23 & 1400 & (-90, 90) & (1.8, 1.4) & ($1.053\times10^{-3}$, $8.186\times10^{-4}$) \\
13+31 & 900 & (-240, 240) & 2.5 & $1.175\times10^{-3}$ \\
12+23+31 & 1600 & 0 & 1.4 & $1.022\times10^{-1}$ \\
13+32+21 & 2000 & 0 & 1.2 & $2.114\times10^{-1}$ \\
All six links & 1800 & 0 & 0.7 & $7.229\times10^{-2}$ \\
\end{tabular}
\end{ruledtabular}
\end{table*}

\begin{figure*}[t]
\centering
\includegraphics[width=0.96\textwidth]{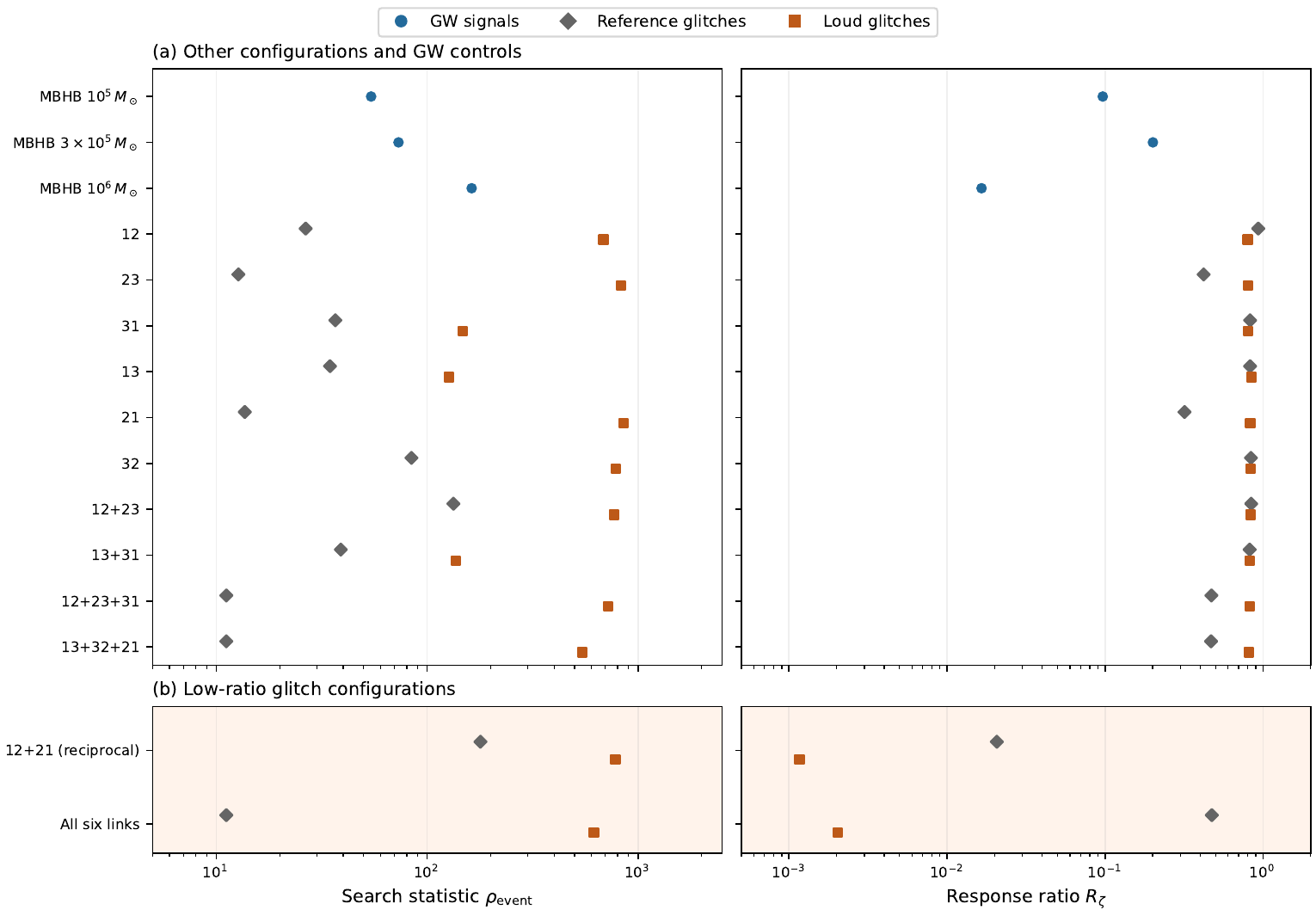}
\caption{Single- and multi-link response comparisons at reference and loud
glitch amplitudes. Circles show GW controls, diamonds reference glitches,
and squares loud glitches. Both columns report the highest-ranked candidate
in each separate injection, using logarithmic axes with shared
limits between the upper and lower panels. (a) Other configurations and
MBHB controls. (b) The reciprocal pair and simultaneous all-six-link case,
which demonstrate a limitation of $R_\zeta$ alone: their loud injections
produce small ratios despite an instrumental origin. The reference
six-link trial's leading candidate has a time-bin-center interval that does
not overlap an injected glitch interval, so its ratio must not be interpreted
as the injected glitch's response. The reference cyclic-triplet candidates
in (a) have the same qualification; the link-21
timing offset is discussed in the text. Parameters are given in
Table~\ref{tab:multilink-injections}. Ratios use background estimates from
the observed data. Plotting and validation code was developed with
assistance from OpenAI Codex (GPT-6 Astra).}
\label{fig:multilink-current}
\end{figure*}

Figure~\ref{fig:multilink-current} retains every configuration alongside the three MBHB control injections. Both reference and loud-glitch trials are shown, each in a separate source-plus-noise record using the same noise realization. The reciprocal pair and all-six-link configuration are placed in a separate lower panel to make the classification limitation explicit. As in the earlier comparisons, timing offsets are discussed separately:
for the link-21 glitch, the leading candidate begins about 558 s after
the input disturbance ends.

Across the primary, generic-burst, and multi-link studies, the loud-injection
leading candidates in fifteen of the seventeen glitch configurations have $R_\zeta=0.800$--$0.845$.
The simultaneous reciprocal pair and six-link disturbance instead give
$1.17\times10^{-3}$ and $2.04\times10^{-3}$, respectively.
These two cases, separated in the lower panel of
Fig.~\ref{fig:multilink-current}, are explicit limitations of glitch classification using $R_\zeta$ alone. The reciprocal-pair example illustrates how correlated instrumental disturbances can suppress the zeta response. The disturbances in readouts 12 and 21 have identical amplitudes and durations and are injected simultaneously. Their small $R_\zeta$ is consistent with cancellation between the oppositely signed, delayed contributions to the zeta combinations, while a response remains in the Sagnac channels used in the denominator. In the equal-arm, constant-delay limit, identical simultaneous reciprocal disturbances can cancel in the zeta combinations. The 13+31 configuration is also a reciprocal pair, but its components are separated by 480 s and have different durations from the 12+21 example. Its larger response ratio is consistent with weaker cancellation, but these tests do not isolate the effects of timing, duration or arm identity.

The six-link example applies the same disturbance simultaneously to all six
directed readouts. This correlated instrumental pattern also gives a small
zeta excess relative to the Sagnac excess on the selected pixels. The ratio
alone therefore cannot distinguish such a disturbance from a GW candidate
with a similarly small ratio. These prescribed examples establish the
response ambiguity, not how often these instrumental patterns will occur.

Thus a small $R_\zeta$ alone cannot exclude an instrumental origin.
The three MBHB leading-candidate ratios are $0.0967$, $0.201$ and $0.0165$
in mass order, and the four generic GW bursts span
$7.25\times10^{-4}$--$0.0272$. Their contrast with the other tested glitch
configurations motivates retaining the ratio alongside the search statistic
for follow-up studies of candidate origin.

\subsection{Reconstruction of simulated MBHB signals}
\label{subsec:simulated-reconstruction}

\begin{figure*}[!t]
\centering
\includegraphics[width=0.94\textwidth]{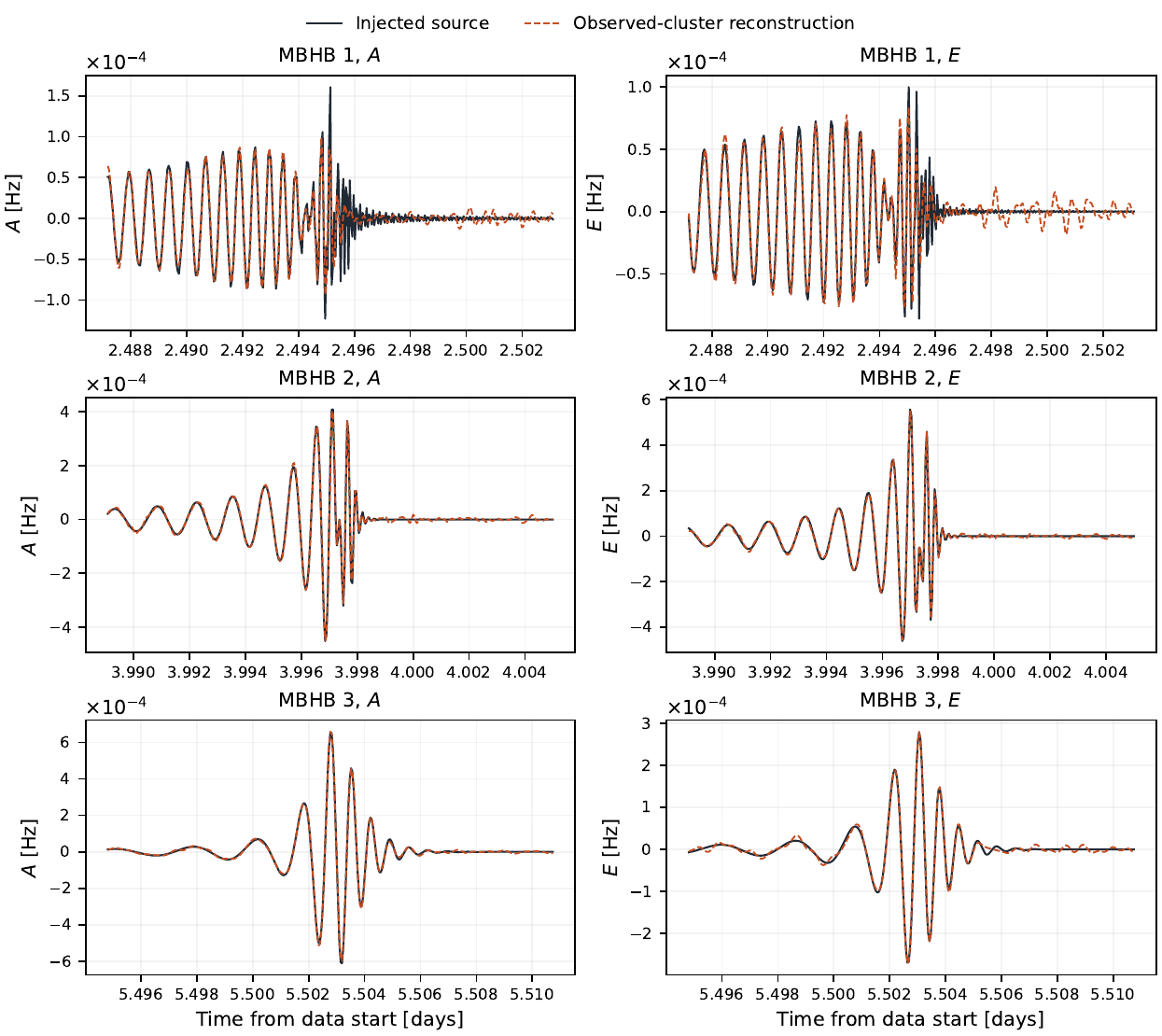}
\caption{Reconstruction of the three simulated MBHBs from raw observed-cluster
A/E coefficients, compared with the injected source-only responses.  Rows are
ordered by increasing mass; columns show A and E.  Each panel covers
$\pm0.008$ day around the peak combined source A/E power to resolve the
waveform. The comparison uses the original common time coordinate and amplitude, with no
fitted rescaling or shift.  Metrics in the text use the full time series,
not only these displayed intervals.  This separate reconstruction study uses
known injection windows; it is not a reconstruction of the Sangria candidates.
Plotting code was revised with assistance from OpenAI Codex
(GPT-6 Astra).}
\label{fig:observed-reconstruction}
\end{figure*}

We now turn from candidate identification and response comparison to waveform recovery.  A separate study reconstructs the three MBHB responses from the noisy
observed A/E data.  It uses $M=512$, $K=256$, the same order and precision parameters, and a clustering band $10^{-5}$--$10^{-1}\,\mathrm{Hz}$. Clustering is performed anew in windows extending $\pm0.75$ days around the response peaks supplied by the injection catalog. Candidate selection preferentially uses a peak within $\pm0.35$ days of that known peak, or overlap with the catalog injection interval. These choices make Fig.~\ref{fig:observed-reconstruction} an injection-informed reconstruction study, separate from blind candidate finding.

The precessing and Sangria follow-ups use a different selection: the original search-selected candidate pixels are retained and used as starting points. We extend their time-bin center interval by two bins on each side and retain the original analysis frequency band. WDM power is recomputed with $K=256$ on the same coefficient grid. Each channel is normalized by its frequency-wise median power over the valid time bins of the full record or chunk. The growth threshold is the 95th percentile of the combined normalized A/E power within the analysis band and the guarded record, not within the local reconstruction window. For the fourteen-day precessing records, this threshold uses days 0.5--13.5; for Sangria it uses the original chunk after half-day guards. Within the enlarged window, pixels at or above this threshold are added if they connect to the retained set, iterating until no further pixels qualify. Connectivity is unchanged from the search. 
This is an empirical starting choice, not an optimized window or a held-out population test. Neither injected waveforms nor fitted source parameters enter this growth rule. Disconnected bright pixels are not merged into the event. 

We reconstruct each observed channel using the two WDM
quadratures employed in cWB's dual-stream
analysis~\cite{Klimenko2016TransientMethod}, implemented
through the inverse transforms provided by
\texttt{wdm-wavelet}~\cite{WDMWaveletPackage}.
With $K=256$, we retain the unwhitened coefficients
within the reconstruction pixel set, then set the remaining
coefficients to zero, and average the time-domain
reconstructions from the two quadratures. This preserves
the channel's physical units.

Figure~\ref{fig:observed-reconstruction} compares the injection-informed aligned-spin reconstruction with the injected source response, without an amplitude fit or a time shift.  For the concatenated A/E source and reconstructed time series $u$ and $v$,
the normalized zero-lag overlap is
$\mathcal{O}=(u\cdot v)/(\lVert u\rVert\lVert v\rVert)$,
the reconstruction-to-source norm ratio is $r=\lVert v\rVert/\lVert u\rVert$,
and the relative residual norm is $\epsilon=\lVert v-u\rVert/\lVert u\rVert$.  The normalized overlaps without a time shift are 0.9315,
0.9934, and 0.9915 in increasing MBHB mass order. The corresponding relative residual norms are 0.3720, 0.1151, and 0.1309; the reconstruction-to-source norm ratios are 1.0087, 0.9967, and 1.0052. The largest discrepancy occurs for the lowest-mass example. Expanding the mask by one frequency bin and one time bin reduces the overlaps to 0.9057, 0.9877, and 0.9886, respectively, so this dilation does not improve these reconstructions. This check does not establish a general window-selection rule for longer signals. The search-seeded follow-ups described above are evaluated separately for precessing signals and for Sangria in Secs.~\ref{subsec:precessing-validation} and \ref{subsec:sangria-targeted-reconstruction}.

\subsection{Precessing MBHB injections and targeted follow-up}
\label{subsec:precessing-validation}

The next three examples connect search-selected candidates to targeted
waveform extraction for signals containing precession and higher harmonics.
The selection is made by the reference search before examining the injected waveform.  We use IMRPhenomTPHM~\cite{Estelles2022IMRPhenomTPHM, GarciaQuirosTiwariBabak2025PhenomXPY}, implemented in
\texttt{phenomxpy}, with detector-frame total masses
$(10^6,3\times10^6,10^7)M_\odot$, common symmetric mass ratio
$\eta=m_1m_2/(m_1+m_2)^2=0.18$, and
$D_L=6230\,\mathrm{Mpc}$.  At the reference frequency
$f_{\rm ref}=2\times10^{-4}\,\mathrm{Hz}$, the dimensionless spin vectors, with components rounded here, are $\boldsymbol{\chi}_1=(0.68,0.14,0.4)$ and $\boldsymbol{\chi}_2=(-0.37,0.42,0.22)$ in the reference source frame, whose $z$ axis is the orbital-angular-momentum direction. Their
magnitudes are 0.8 and 0.6.  The inclination to this reference axis is $1.35\,\mathrm{rad}$ and the reference phase is zero.  The model is initialized
with $f_{\rm min}=10^{-4}\,\mathrm{Hz}$ for the first two masses and
$5\times10^{-5}\,\mathrm{Hz}$ for $10^7M_\odot$, with the coprecessing modes
$(2,\pm2)$, $(2,\pm1)$, $(3,\pm3)$, $(4,\pm4)$, and $(5,\pm5)$.

Each source is propagated through the existing link-response and second-generation TDI chain and added separately to the same simulated 14-day instrument noise realization.  In mass order, the sky coordinates supplied as right ascension and declination are $(0.45,0.35)$, $(1.55,-0.40)$, and $(3.10,0.12)\,\mathrm{rad}$; the combined polarization-amplitude peaks are placed at days 2.15, 6.85, and 10.95.  All three inspirals extend before the observing interval, so only a finite observation is evaluated. Their strain inputs extend 1200 s before the link-output start and after the
simulated data to cover retarded response times. For $10^7M_\odot$, the lower starting frequency places model
initialization about 27 days before its time origin, outside the observation. The original TDI frequency cut is retained, so comparisons refer to the band-passed source contribution.

\begin{table}[!htbp]
\caption{Highest-ranked candidates in the three precessing-source searches.
$\mathcal{O}$, $r$, and $\epsilon$ denote the normalized zero-lag overlap,
reconstruction/source norm ratio, and relative residual for concatenated A/E
samples within the enlarged reconstruction window.  The source
reference is the band-passed contribution. The highest-mass row
uses the earlier-start waveform.}
\label{tab:precessing-reconstruction}
\begin{ruledtabular}
\begin{tabular}{ccccc}
$M_{\rm det}/(10^6M_\odot)$ & $\rho_{\rm event}$ & $\mathcal{O}$ & $r$ &
$\epsilon$ \\
\hline
1 & 47.05 & 0.9653 & 0.9594 & 0.2612 \\
3 & 438.28 & 0.9950 & 0.9925 & 0.0998 \\
10 & 123.10 & 0.9077 & 1.1034 & 0.4631 \\
\end{tabular}
\end{ruledtabular}
\end{table}

The search uses the same WDM $M=512$, $K=32$ A/E
search over days 0.5--13.5, with the $10^{-5}$--$0.02\,\mathrm{Hz}$ band. All candidates are retained for inspection;
Table~\ref{tab:precessing-reconstruction}
reports the highest-ranked candidates, which are associated with the injected MBHBs. For reconstruction, the observed data are transformed again and the search-selected pixels seed the connected growth defined in Sec.~\ref{subsec:simulated-reconstruction}. The original search candidates and ranking statistics are unchanged.

\begin{figure*}[!t]
\centering
\includegraphics[width=0.98\textwidth]
  {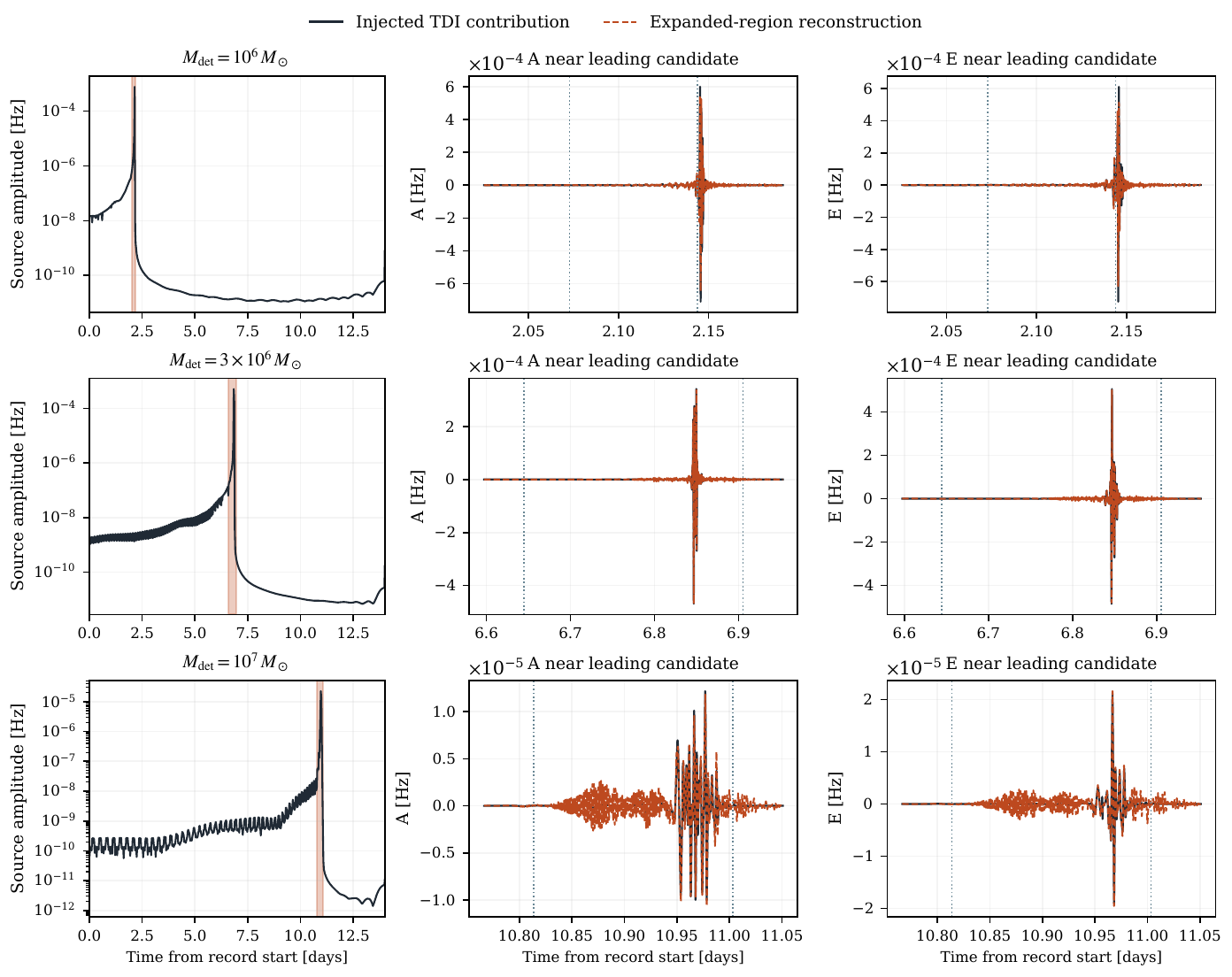}
\caption{Precessing sources and their targeted reconstructions, in increasing
mass from top to bottom.  Left: maximum of $(A^2+E^2)^{1/2}$ for the injected
TDI contribution in consecutive 2048-s bins over the entire data stretch; shading
marks the enlarged reconstruction window.  All three sources include an ongoing inspiral before the observation starts.  Center and right: raw source contributions and unfitted reconstructions near the leading candidate over the enlarged reconstruction window; dotted lines mark the original candidate-center interval. TDI amplitudes are in Hz. Plotting code was revised with assistance from OpenAI Codex (GPT-6 Astra).}
\label{fig:precessing-reconstruction}
\end{figure*}

Figure~\ref{fig:precessing-reconstruction} illustrates the
reconstruction of three precessing massive-black-hole binaries. Together with the aligned-spin examples, these results demonstrate that the pipeline can recover signals with more complex waveform structure without imposing a waveform template. The reconstructed channel responses follow the main features of the injected signals, although the agreement varies across the three examples, with larger deviations in the highest-mass case. Table~\ref{tab:precessing-reconstruction} quantifies this agreement. These examples extend the demonstrated reconstruction capability to the precessing binaries considered here.

\subsection{Sangria TDI-data analysis}
\label{subsec:sangria-results}

\begin{figure*}[!t]
\centering
\includegraphics[width=0.94\textwidth]{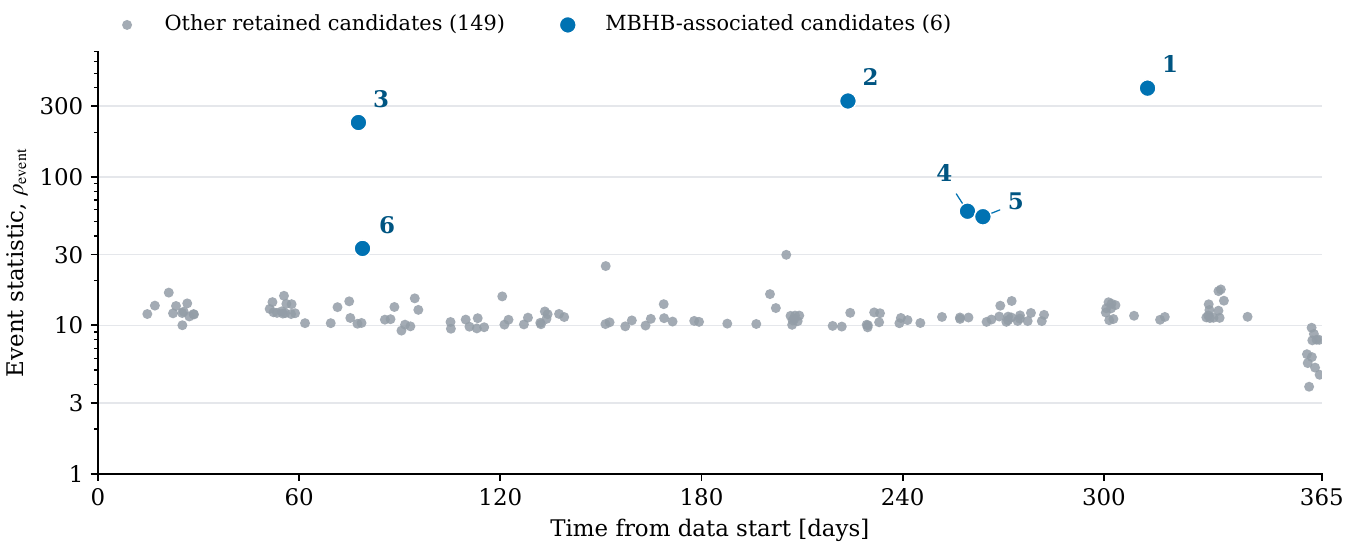}
\caption{
Retained Sangria candidates across the one-year data set. Each dot shows a search
candidate's peak time and event statistic $\rho_{\rm event}$, with a
logarithmic vertical axis. All 155 retained candidates are shown; the search
keeps at most 12 per chunk and excludes half-day chunk edges, as described in
the text. Blue dots
highlight the six highest-ranked candidates associated with injected MBHBs
using the unblinded catalog and MBHB-only TDI contribution; labels give their
ranks in Table~\ref{tab:sangria-top-ten}.  Gray dots denote the remaining
149 candidates, which are not classified in this comparison.  The highlighted
points represent source associations. Plotting code was developed with assistance from OpenAI Codex (GPT-6 Astra).}
\label{fig:sangria-demo}
\end{figure*}

The final study applies the same candidate-finding approach to the Sangria mock data challenge, then uses the unblinded information to investigate the leading candidates. The A/E WDM search is applied to the public Sangria LISA Data Challenge data set, which provides one year of sampled \XYZ{} TDI data in dimensionless fractional-frequency units at $\Delta t=5\,{\rm s}$~\cite{SangriaLDC2a2022}.  The pipeline
converts \XYZ{} to \AET{} using the linear combinations in Sec.~\ref{subsec:tdi-observables} and applies the same self-A/E WDM analysis to nonoverlapping 30-day chunks, with a shorter final chunk. The search excludes half a day at each chunk edge, giving approximately 352 days of nominal searched windows within the 365-day record. At most the 12 highest-ranked candidates are retained from each chunk. The resulting catalog contains 155 retained candidates, rather than every candidate passing the pixel and size cuts. Because the Sangria product used here provides TDI observables rather than raw one-way links, this analysis is restricted to the
\Achan{}/\Echan{} WDM analysis and not the glitch classification.

The retained catalog is ranked across chunks; the three loudest candidates peak at days 313.016, 223.667, and 77.695 with $\rho_{\rm event}=396.44$, 324.85, and 232.61, respectively. The ten loudest candidates are listed in Table~\ref{tab:sangria-top-ten}. Figure~\ref{fig:sangria-demo} shows the event statistics of all retained candidates across the year. It highlights the six highest-ranked
candidates, which we associate with the six injected MBHBs using the unblinded Sangria data.

The six associations are distinct in coalescence time
(Table~\ref{tab:sangria-top-ten}).  Direct comparison of the observed \Achan{}/\Echan{} time series with the aggregate MBHB-only TDI contribution confirms the injected transients at these epochs, without fitting an amplitude or shifting the time series. Candidate peak times differ from catalog
coalescence by $-498$ to $+796\,{\rm s}$.  These time-frequency peak-bin centers should not be understood as the estimates of coalescence time. 

\begin{table}[!htbp]
\caption{Ten highest-ranked retained Sangria A/E WDM candidates.
$t_p$ and $f_p$ are the time and frequency of the pixel with largest
$P_{\rm AE}$ in each candidate; $N_{\rm pix}$ is the number of selected
pixels. $\rho_{\rm event}$ and $C_{\rm AE}$ are defined in
Eqs.~(\ref{eq:rho-event}) and (\ref{eq:ae-coherence}).
The last column gives the catalog coalescence time $t_c$ for the six MBHB
associations; dashes indicate candidates not classified in this comparison.}
\label{tab:sangria-top-ten}
\scriptsize
\setlength{\tabcolsep}{2.2pt}
\renewcommand{\arraystretch}{0.92}
\begin{ruledtabular}
\begin{tabular}{ccccccc}
Rank & $t_p$ [d] & $f_p$ [mHz] & $N_{\rm pix}$ & $\rho_{\rm event}$ & $C_{\rm
AE}$ & $t_c$ [d] \\
\hline
1 & 313.016 & 3.125 & 45 & 396.44 & 0.999 & 313.011 \\
2 & 223.667 & 4.687 & 49 & 324.85 & 0.996 & 223.658 \\
3 & 77.695 & 5.859 & 63 & 232.61 & 0.997 & 77.690 \\
4 & 259.324 & 2.344 & 54 & 58.70 & 0.927 & 259.328 \\
5 & 263.914 & 2.148 & 32 & 53.93 & 0.977 & 263.905 \\
6 & 78.909 & 1.563 & 34 & 32.96 & 0.805 & 78.915 \\
7 & 205.276 & 2.734 & 212 & 29.86 & 0.778 & --- \\
8 & 151.436 & 0.781 & 235 & 25.07 & 0.925 & --- \\
9 & 334.872 & 5.078 & 35 & 17.45 & 0.135 & --- \\
10 & 334.161 & 9.961 & 39 & 17.00 & 0.615 & --- \\
\end{tabular}
\end{ruledtabular}
\end{table}

\subsubsection{Targeted reconstruction following the search}
\label{subsec:sangria-targeted-reconstruction}

\begin{table}[b]
\caption{Local targeted Sangria reconstruction checks. $u$ is the aggregate
injected MBHB A/E contribution and $v$ the unfitted candidate reconstruction,
restricted to the enlarged reconstruction window.  $\mathcal{O}$ is their
normalized zero-lag overlap, $r=\lVert v\rVert/\lVert u\rVert$, and
$\epsilon=\lVert v-u\rVert/\lVert u\rVert$.}
\label{tab:sangria-targeted-reconstruction}
\begin{ruledtabular}
\begin{tabular}{cccc}
Rank & $\mathcal{O}$ & $r$ & $\epsilon$ \\
\hline
1 & 0.9956 & 1.0044 & 0.0939 \\
2 & 0.9989 & 0.9990 & 0.0477 \\
3 & 0.9940 & 1.0009 & 0.1092 \\
4 & 0.9912 & 1.0123 & 0.1341 \\
5 & 0.9963 & 0.9997 & 0.0865 \\
6 & 0.9940 & 0.9952 & 0.1092 \\
\end{tabular}
\end{ruledtabular}
\end{table}

\begin{figure*}[!t]
\centering
\includegraphics[width=0.90\textwidth]{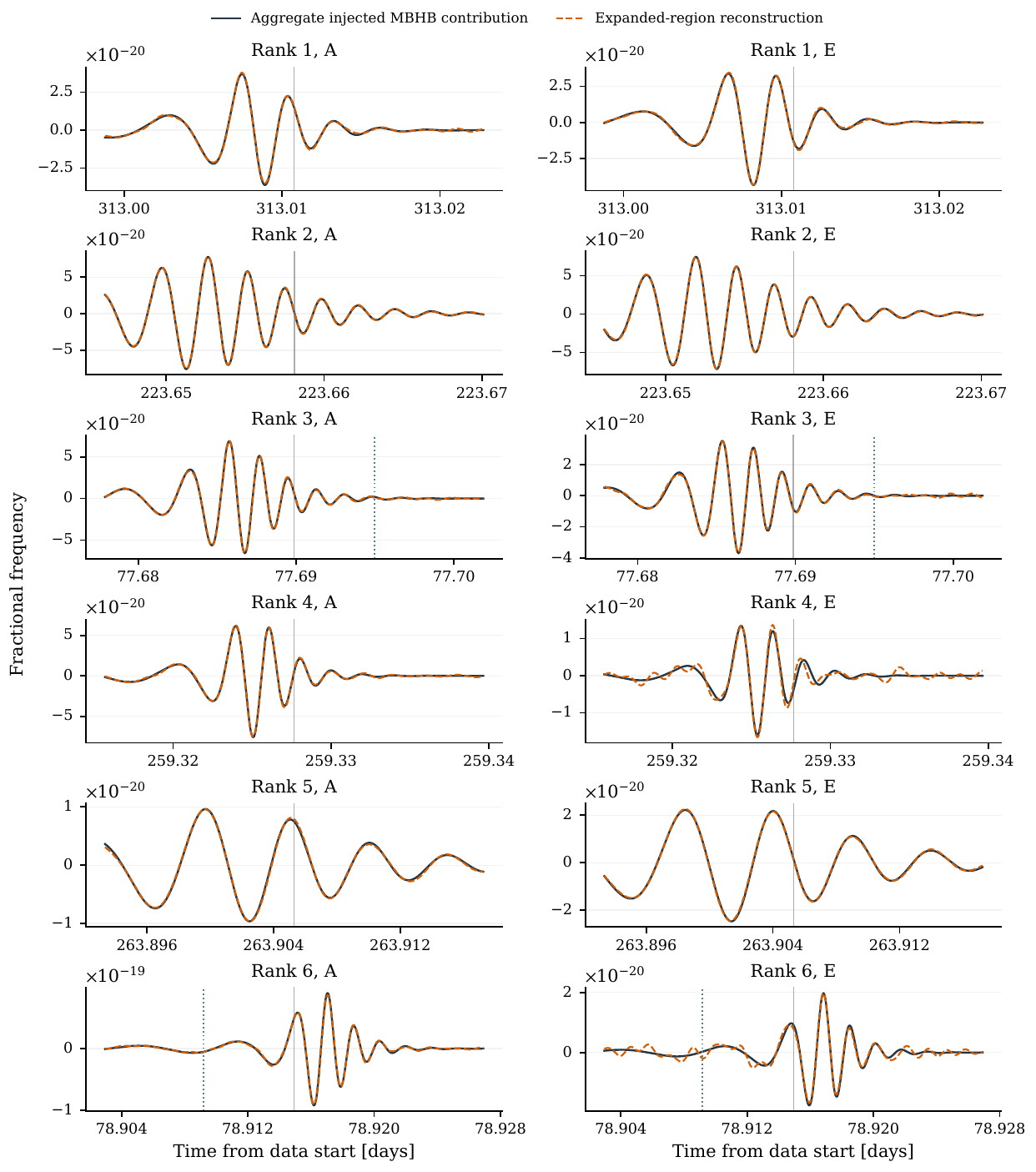}
\caption{Waveform reconstruction for the six leading Sangria candidates, ordered
by rank from top to bottom; columns show the two measured data streams.
The search-selected regions seed connected growth within an enlarged
reconstruction window, using the procedure described in
Sec.~\ref{subsec:simulated-reconstruction}. The recovered signal is
compared with the combined contribution of the injected black-hole binaries,
without fitted rescaling or
shifting.  Each panel shows $\pm0.012$ day around the catalog coalescence time
(solid vertical line), solely to display the merger region on the original
time coordinate.  Dotted vertical lines mark original candidate-center
boundaries when visible. Table~\ref{tab:sangria-targeted-reconstruction}
uses the enlarged reconstruction
windows, not these narrow display windows.  A/E amplitudes are dimensionless
fractional-frequency TDI.  The expanded selection now recovers the merger of
rank 6 beyond its original candidate-center interval.  The reference combines
all injected black-hole signals; isolated source time series are not supplied.
Plotting code was developed and revised with assistance from OpenAI Codex
(GPT-6 Astra).}
\label{fig:sangria-targeted-combined}
\end{figure*}

The six leading candidates are followed up with a separate $K=256$ WDM reconstruction, retaining $M=512$, order 6, precision 6, and the native $5\,\mathrm{s}$ cadence.  The original $K=32$ candidate masks and ranking statistics are first reproduced on their original 30-day chunks.  The $K=256$ coefficient grids have identical time and frequency centers, allowing the original pixel-index masks to seed connected growth in the new observed A/E maps. The two-bin time extension, connected growth above the fixed 95th-percentile threshold, and normalization follow Sec.~\ref{subsec:simulated-reconstruction}; the band
remains $10^{-5}$ -- $0.02\,\mathrm{Hz}$ and the percentile is calculated over the original chunk with half-day guards.


Figure~\ref{fig:sangria-targeted-combined} compares the unfitted candidate reconstructions with the aggregate MBHB
contribution.  The reference curves are obtained from the combined MBHB-only TDI signal supplied with the unblinded Sangria data, transformed into the same A and E channels as the observations using the linear combinations in Sec.~\ref{subsec:tdi-observables}. The unblinded data do not provide isolated TDI time series for each source.  Table~\ref{tab:sangria-targeted-reconstruction} therefore reports local comparisons on the enlarged reconstruction
windows.  Each metric concatenates A and E and uses the time-domain definitions in Sec.~\ref{subsec:simulated-reconstruction}.

The reconstructions recover the main waveform features of all six MBHB events identified in the Sangria data, with overlaps of 0.9912--0.9989 against the aggregate injected MBHB contribution in the corresponding comparison windows. The reconstructed norms are within approximately 1.3\% of the reference values, indicating close agreement in overall amplitude. Small residual differences remain, with relative residuals of 0.0477--0.1341. These may partly reflect noise admitted by additional selected pixels, although this contribution has not been isolated. The results demonstrate
accurate local waveform recovery for all six events.

\section{Conclusions}
\label{sec:conclusions}

We have presented \pversion{}, a pipeline that combines an unmodeled
transient search, waveform reconstruction, and a glitch classifier for
LISA data. The search identifies concentrations of power in time and
frequency without requiring a source waveform template. We tested it
with simulated black-hole mergers and burst signals passed through the
LISA response, together with instrumental glitches injected into the
readouts. The glitch classifier compares candidate responses across
different combinations of the measurements, using a background estimated
from the observed data. Most glitch configurations tested produce larger
response ratios than the gravitational-wave examples. The tested
simultaneous disturbances in a reciprocal pair of links and in all six
links also produce small ratios. The response ratio therefore provides
information about a candidate's origin, but cannot resolve these cases
without additional evidence.

The reconstruction studies recover waveform structure from both
aligned-spin and precessing binaries. Agreement varies among the
precessing examples, with larger deviations in the highest-mass case.
In the unblinded Sangria analysis, all six injected massive-black-hole
binaries are recovered as the six highest-ranked candidates. Their local
reconstructions recover the main waveform features, with normalized
overlaps of 0.9912--0.9989. These overlaps measure agreement in waveform
shape against the sum of the injected black-hole signals supplied with
the unblinded data. The comparisons therefore assess reconstruction over
the intervals studied using this combined reference.

These results motivate a search for LISA transients whose detailed
waveforms are uncertain. Precession, eccentricity, and other physical
effects can introduce features beyond those described by a restricted
waveform family. The global-fit demonstrations cited here use
aligned-spin, quasi-circular massive-black-hole
waveforms~\cite{LittenbergCornish2023GLASS,Katz2025Erebor,Deng2025ModularGlobalFit}.
If additional effects substantially change the signal, a search based on
that restricted family can lose sensitivity or leave unexplained
residuals. The precessing examples studied here show how such signals can
be investigated without source templates; eccentric binaries remain a
target for future validation. The burst studies demonstrate recovery of
several different signal shapes without assigning them to a particular
astrophysical population.

A candidate list from \pversion{} could help identify black-hole mergers,
possible glitches, and other transients before an initial search for
Galactic binaries. Candidate reconstructions could provide starting
estimates of the transient contributions for the iterative analysis used
in global-fit
pipelines~\cite{LittenbergCornish2023GLASS,Katz2025Erebor,Deng2025ModularGlobalFit}.
Repeating the search on the residual data could also reveal features
missed by the source models. Rapid identification of interesting
intervals could guide low-latency follow-up before a complete analysis
of the overlapping sources. Testing these applications within the
relevant pipelines can establish whether they improve the subsequent
recovery of other signals.

The search can also be adapted to particular scientific targets. Its
present preference for increasing frequency is an empirical choice for
the examples considered here. This choice can be revisited together with
the rules for grouping signal power and selecting the reconstruction
region. Approximate descriptions of frequency tracks and their harmonics
could help collect signal power that the present grouping rules leave
separated, for example in searches for extreme-mass-ratio inspirals or
eccentric binaries. Such adaptations would introduce assumptions about
the signal's time-frequency structure, but could improve sensitivity
without requiring a detailed prediction of its phase throughout the
observation. Searches for long-lived sources would also need to combine
information over longer observing intervals.

Waveform reconstruction offers a further way to investigate source
physics. For GW150914, cWB reconstructions were compared with
numerical-relativity waveforms to assess consistency with a
binary-black-hole signal~\cite{Abbott2016GW150914Minimal}. Similar
comparisons in LISA could reveal signal structure that a source model
does not explain, guiding studies of unexpected dynamics, environmental
influences, or departures from general relativity. This approach is
particularly useful when reliable templates for the additional physics
are unavailable. A discrepancy would motivate further investigation
without identifying a unique cause. Environmental effects and
eccentricity can themselves mimic departures from general relativity
when omitted from the source
model~\cite{Garg2024LISASystematics}. Instrumental effects and waveform
inaccuracies must also be examined before attributing an unexplained
feature to new physics.

Future development should extend the background estimation and response
studies to data with changing noise levels, unresolved foregrounds, and
a wider range of instrumental glitches. Work on time-varying noise and
correlations between wavelet coefficients can guide these
studies~\cite{Cornish2025WaveletCovariance,Cornish2026DynamicNoise,Johnson2026WDMCovariance}.
Larger injection studies can measure the fraction of sources recovered
across a population, while background studies can relate the candidate
ranking to false-alarm probabilities. These developments would strengthen
the pipeline for both targeted source studies and searches for transient
signals beyond the waveform families included in an analysis.
\begin{acknowledgments}
The authors thank Johan Robertsson for fruitful discussions and continued support. 

The analysis uses
PyTDI, LISA Orbits, LISA GW Response, LISA Instrument, \texttt{wdm-wavelet},
\texttt{phenomxpy}, and \texttt{lisaglitch}. OpenAI Codex (GPT-6 Astra) assisted with the literature and citation checks under the authors' direction. All the other scripts and plots are made by authors' unless otherwise specified. 

S.T. is partially supported by the Tomalla Foundation.

Y.X. was supported by the INVESTIGA@UIB programme of the Universitat de les Illes Balears (UIB), co-funded by the 2023 Sustainable Tourism Promotion Plan (ITS2023-086 -- Research Promotion Programme).
This work was supported by the Universitat de les Illes Balears (UIB) with funds from the Programa de Foment de la Recerca i la Innovaci\'{o} de la UIB 2024-2026 (supported by the yearly plan of the Tourist Stay Tax ITS2023-086); the Spanish Agencia Estatal de Investigaci\'{o}n grants PID2022-138626NB-I00, RED2024-153978-E, RED2024-153735-E, funded by MICIU/AEI/10.13039/501100011033 and the ERDF/EU; and the Comunitat Aut\`{o}noma de les Illes Balears through an "Ajut per a projectes de recerca científica i tecnològica (PRD2025\_[099/100]) and through the Conselleria d'Educaci\'{o} i Universitats with funds from the European Union - European Regional Development Fund (ERDF) (SINCO2022/18146 - Plataforma HiTech-IAC3-BIO).

\end{acknowledgments}

\section*{Data Availability}
The Sangria challenge data are available from the LISA Data Challenge
release~\cite{SangriaLDC2a2022}.  The algorithm and supporting data for the
present study will be made publicly available.

\bibliography{cwb_space_prx}

@misc{LISADefinitionStudy2024,
  author = {Colpi, Monica and others},
  title = {{LISA} Definition Study Report},
  year = {2024},
  eprint = {2402.07571},
  archivePrefix = {arXiv},
  primaryClass = {astro-ph.CO},
  url = {https://arxiv.org/abs/2402.07571},
  note = {{ESA-SCI-DIR-RP-002} (September 2023); arXiv release February 2024}
}

@misc{SGSConventions2026,
  author = {Baghi, Quentin and Babak, Stanislas and Barack, Leor and Bayle,
    Jean-Baptiste and Burke, Ollie and Enficiaud, Raffi and Estelles, Hector and
    Garc{\'i}a Quir{\'o}s, Cecilio and Hartwig, Olaf and Hees, Aurelien and
    Husa, Sascha and Inchausp{\'e}, Henri and Joffre, Eric and Klein, Antoine
    and Lynch, Philip and Marsat, Sylvain and Menu, Jonathan and Nasipak, Zach
    and Pardo De Santayana, Ramon and Pfeiffer, Harald and Pound, Adam and
    Pratten, Geraint and Ramos-Buades, Antoni and Sopuerta, Carlos and
    Warburton, Niels},
  title = {{LISA} science ground segment conventions},
  year = {2026},
  eprint = {2603.22377},
  archivePrefix = {arXiv},
  primaryClass = {astro-ph.IM},
  url = {https://arxiv.org/abs/2603.22377},
  note = {Report LISA-DDPC-SEG-TN-007}
}

@article{Tinto2005TDI,
  author = {Tinto, Massimo and Dhurandhar, Sanjeev V.},
  title = {Time-Delay Interferometry},
  journal = {Living Reviews in Relativity},
  volume = {8},
  pages = {4},
  year = {2005},
  doi = {10.12942/lrr-2005-4},
  eprint = {gr-qc/0409034},
  archivePrefix = {arXiv}
}

@article{Prince2002OptimalSensitivity,
  author = {Prince, T. A. and Tinto, M. and Larson, S. L. and Armstrong, J. W.},
  title = {{LISA} optimal sensitivity},
  journal = {Phys. Rev. D},
  volume = {66},
  pages = {122002},
  year = {2002},
  doi = {10.1103/PhysRevD.66.122002},
  eprint = {gr-qc/0209039},
  archivePrefix = {arXiv}
}

@article{Dhurandhar2002AlgebraicTDI,
  author = {Dhurandhar, Sanjeev V. and Nayak, K. Rajesh and Vinet, Jean-Yves},
  title = {Algebraic approach to time-delay data analysis for {LISA}},
  journal = {Phys. Rev. D},
  volume = {65},
  pages = {102002},
  year = {2002},
  doi = {10.1103/PhysRevD.65.102002},
  eprint = {gr-qc/0112059},
  archivePrefix = {arXiv}
}

@article{Muratore2022NoiseMonitors,
  author = {Muratore, Martina and Vetrugno, Daniele and Vitale, Stefano and
    Hartwig, Olaf},
  title = {Time delay interferometry combinations as instrument noise monitors
    for {LISA}},
  journal = {Phys. Rev. D},
  volume = {105},
  pages = {023009},
  year = {2022},
  doi = {10.1103/PhysRevD.105.023009},
  eprint = {2108.02738},
  archivePrefix = {arXiv},
  primaryClass = {gr-qc}
}

@article{HartwigMuratore2022TDI,
  author = {Hartwig, Olaf and Muratore, Martina},
  title = {Characterization of time delay interferometry combinations for the
    {LISA} instrument noise},
  journal = {Phys. Rev. D},
  volume = {105},
  pages = {062006},
  year = {2022},
  doi = {10.1103/PhysRevD.105.062006},
  eprint = {2111.00975},
  archivePrefix = {arXiv},
  primaryClass = {gr-qc}
}

@article{TintoDhurandharMalakar2023G2TDI,
  author = {Tinto, Massimo and Dhurandhar, Sanjeev and Malakar, Dishari},
  title = {Second-generation time-delay interferometry},
  journal = {Phys. Rev. D},
  volume = {107},
  pages = {082001},
  year = {2023},
  doi = {10.1103/PhysRevD.107.082001},
  eprint = {2212.05967},
  archivePrefix = {arXiv},
  primaryClass = {gr-qc}
}

@article{BayleHartwig2023Instrument,
  author = {Bayle, Jean-Baptiste and Hartwig, Olaf},
  title = {Unified model for the {LISA} measurements and instrument
    simulations},
  journal = {Phys. Rev. D},
  volume = {107},
  pages = {083019},
  year = {2023},
  doi = {10.1103/PhysRevD.107.083019},
  eprint = {2212.05351},
  archivePrefix = {arXiv},
  primaryClass = {gr-qc}
}

@article{GarciaQuirosTiwariBabak2025PhenomXPY,
  author = {Garc{\'i}a-Quir{\'o}s, Cecilio and Tiwari, Shubhanshu and Babak,
    Stanislav},
  title = {{GPU}-accelerated {LISA} parameter estimation with full time-domain
    response},
  journal = {Phys. Rev. D},
  volume = {112},
  pages = {064017},
  year = {2025},
  doi = {10.1103/79kn-53nt},
  eprint = {2501.08261},
  archivePrefix = {arXiv},
  primaryClass = {gr-qc},
  url = {https://arxiv.org/abs/2501.08261}
}

@misc{PhenomXPY,
  author = {Garc{\'i}a-Quir{\'o}s, Cecilio},
  title = {{phenomxpy}},
  url = {https://gitlab.com/imrphenom-dev/phenomxpy},
  note = {Accessed 19 October 2026},
  year = {n.d.}
}

@article{Necula2012FastWDM,
  author = {Necula, V. and Klimenko, S. and Mitselmakher, G.},
  title = {Transient analysis with fast {Wilson-Daubechies} time-frequency
    transform},
  journal = {Journal of Physics: Conference Series},
  volume = {363},
  pages = {012032},
  year = {2012},
  doi = {10.1088/1742-6596/363/1/012032}
}

@article{Klimenko2016TransientMethod,
  author = {Klimenko, S. and Vedovato, G. and Drago, M. and Salemi, F. and
    Tiwari, V. and Prodi, G. A. and Lazzaro, C. and Ackley, K. and Tiwari, S.
    and {Da Silva}, C. F. and Mitselmakher, G.},
  title = {Method for detection and reconstruction of gravitational wave
    transients with networks of advanced detectors},
  journal = {Phys. Rev. D},
  volume = {93},
  pages = {042004},
  year = {2016},
  doi = {10.1103/PhysRevD.93.042004},
  eprint = {1511.05999},
  archivePrefix = {arXiv},
  primaryClass = {gr-qc}
}

@misc{WDMWaveletPackage,
  author = {Xu, Yumeng},
  title = {{wdm-wavelet}: {WDM} wavelet transform},
  url = {https://pypi.org/project/wdm-wavelet/},
  note = {Accessed 19 October 2026},
  year = {n.d.}
}

@misc{Babak2021Sensitivity,
  author = {Babak, Stanislav and Hewitson, Martin and Petiteau, Antoine},
  title = {{LISA} sensitivity and {SNR} calculations},
  year = {2021},
  eprint = {2108.01167v1},
  archivePrefix = {arXiv},
  primaryClass = {astro-ph.IM},
  url = {https://arxiv.org/abs/2108.01167v1}
}

@article{RobsonCornish2019Glitches,
  author = {Robson, Travis and Cornish, Neil J.},
  title = {Detecting gravitational wave bursts with {LISA} in the presence of
    instrumental glitches},
  journal = {Phys. Rev. D},
  volume = {99},
  pages = {024019},
  year = {2019},
  doi = {10.1103/PhysRevD.99.024019},
  eprint = {1811.04490},
  archivePrefix = {arXiv},
  primaryClass = {gr-qc}
}

@article{Baghi2022LPFGlitches,
  author = {Baghi, Quentin and Korsakova, Natalia and Slutsky, Jacob and
    Castelli, Eleonora and Karnesis, Nikolaos and Bayle, Jean-Baptiste},
  title = {Detection and characterization of instrumental transients in {LISA
    Pathfinder} and their projection to {LISA}},
  journal = {Phys. Rev. D},
  volume = {105},
  pages = {042002},
  year = {2022},
  doi = {10.1103/PhysRevD.105.042002},
  eprint = {2112.07490},
  archivePrefix = {arXiv},
  primaryClass = {gr-qc}
}

@article{Spadaro2023GlitchSystematics,
  author = {Spadaro, Alice and Buscicchio, Riccardo and Vetrugno, Daniele and
    Klein, Antoine and Gerosa, Davide and Vitale, Stefano and Dolesi, Rita and
    Weber, William Joseph and Colpi, Monica},
  title = {Glitch systematics on the observation of massive black-hole binaries
    with {LISA}},
  journal = {Phys. Rev. D},
  volume = {108},
  pages = {123029},
  year = {2023},
  doi = {10.1103/PhysRevD.108.123029},
  eprint = {2306.03923},
  archivePrefix = {arXiv},
  primaryClass = {gr-qc}
}

@article{Muratore2025GlitchPipeline,
  author = {Muratore, Martina and Gair, Jonathan and Hartwig, Olaf and Katz,
    Michael L. and Toubiana, Alexandre},
  title = {Pipeline for searching and fitting instrumental glitches in {LISA}
    data},
  year = {2025},
  eprint = {2505.19870},
  archivePrefix = {arXiv},
  primaryClass = {gr-qc},
  url = {https://arxiv.org/abs/2505.19870},
  journal = {Phys. Rev. D},
  volume = {112},
  pages = {063041},
  doi = {10.1103/1sj2-219n}
}

@misc{SangriaLDC2a2022,
  author = {Le Jeune, Maude and Babak, Stanislav},
  title = {{LISA Data Challenge Sangria (LDC2a)}, Version v2},
  year = {2022},
  doi = {10.5281/zenodo.7132178},
  howpublished = {\url{https://doi.org/10.5281/zenodo.7132178}},
  url = {https://lisa-ldc.in2p3.fr/challenge2a},
  note = {LISA Data Challenge data set; description accessed 19 October 2026}
}

@misc{PyTDI,
  author = {Staab, Martin and Bayle, Jean-Baptiste and Hartwig, Olaf},
  title = {{PyTDI}},
  howpublished = {\url{https://pypi.org/project/pytdi/}},
  note = {Accessed 19 October 2026},
  year = {n.d.}
}

@misc{LISAOrbits,
  author = {Bayle, Jean-Baptiste and Hees, Aur{\'e}lien and Lilley, Marc and {Le
    Poncin-Lafitte}, Christophe and Martens, Waldemar and Joffre, Eric},
  title = {{LISA Orbits}},
  howpublished = {\url{https://pypi.org/project/lisaorbits/}},
  note = {Accessed 19 October 2026},
  year = {n.d.}
}

@misc{LISAGWResponse,
  author = {Bayle, Jean-Baptiste and Baghi, Quentin and Renzini, Arianna and {Le
    Jeune}, Maude},
  title = {{LISA GW Response}},
  howpublished = {\url{https://pypi.org/project/lisagwresponse/}},
  note = {Accessed 19 October 2026},
  year = {n.d.}
}

@misc{LISAInstrument,
  author = {Bayle, Jean-Baptiste and Hartwig, Olaf and Kastaun, Wolfgang and
    Staab, Martin},
  title = {{LISA Instrument}},
  howpublished = {\url{https://pypi.org/project/lisainstrument/}},
  note = {Accessed 19 October 2026},
  year = {n.d.}
}

@misc{LISAGlitch,
  author = {Bayle, Jean-Baptiste},
  title = {{LISA Glitch}},
  howpublished = {\url{https://pypi.org/project/lisaglitch/}},
  note = {Accessed 19 October 2026},
  year = {n.d.}
}

@article{Abbott2016GW150914Minimal,
  author = {Abbott, B. P. and others},
  collaboration = {LIGO Scientific Collaboration and Virgo Collaboration},
  title = {Observing gravitational-wave transient {GW150914} with minimal
    assumptions},
  journal = {Phys. Rev. D},
  volume = {93},
  pages = {122004},
  year = {2016},
  doi = {10.1103/PhysRevD.93.122004},
  eprint = {1602.03843},
  archivePrefix = {arXiv},
  primaryClass = {gr-qc},
  note = {Publisher's Note: Phys. Rev. D 94, 069903 (2016)}
}

@article{Abbott2020GW190521,
  author = {Abbott, R. and others},
  collaboration = {LIGO Scientific Collaboration and Virgo Collaboration},
  title = {{GW190521}: A Binary Black Hole Merger with a Total Mass of
    {$150\,M_{\odot}$}},
  journal = {Phys. Rev. Lett.},
  volume = {125},
  pages = {101102},
  year = {2020},
  doi = {10.1103/PhysRevLett.125.101102},
  eprint = {2009.01075},
  archivePrefix = {arXiv},
  primaryClass = {gr-qc}
}

@article{Knee2024LISABursts,
  author = {Knee, Alan M. and McIver, Jess and Naoz, Smadar and Romero-Shaw,
    Isobel M. and Hoang, Bao-Minh and Grishin, Evgeni},
  title = {Detecting Gravitational-wave Bursts from Black Hole Binaries in the
    Galactic Center with {LISA}},
  journal = {The Astrophysical Journal Letters},
  volume = {971},
  pages = {L38},
  year = {2024},
  doi = {10.3847/2041-8213/ad6a10},
  eprint = {2404.12571},
  archivePrefix = {arXiv},
  primaryClass = {astro-ph.HE}
}

@article{GairJones2007HACRPreprint,
  author = {Gair, Jonathan R. and Jones, Gareth},
  title = {Detecting extreme mass ratio inspiral events in {LISA} data using the Hierarchical Algorithm for Clusters and Ridges ({HACR})},
  journal = {Class. Quantum Grav.},
  volume = {24},
  number = {5},
  pages = {1145--1168},
  year = {2007},
  doi = {10.1088/0264-9381/24/5/007},
  eprint = {gr-qc/0610046v2},
  archivePrefix = {arXiv}
}

@misc{NASALISAMissionReference,
  author = {{NASA}},
  title = {{LISA}: Reference Documents},
  url = {https://lisa.nasa.gov/documentsReference.html},
  note = {Accessed 19 October 2026; 2024 LISA Definition Study Report summary}
}

@article{Estelles2022IMRPhenomTHM,
  author = {Estell{\'e}s, H{\'e}ctor and Husa, Sascha and Colleoni, Marta and
    Keitel, David and Mateu-Lucena, Maite and Garc{\'i}a-Quir{\'o}s, Cecilio and
    Ramos-Buades, Antoni and Borchers, Angela},
  title = {Time-domain phenomenological model of gravitational-wave subdominant
    harmonics for quasicircular nonprecessing binary black hole coalescences},
  journal = {Phys. Rev. D},
  volume = {105},
  pages = {084039},
  year = {2022},
  doi = {10.1103/PhysRevD.105.084039},
  eprint = {2012.11923},
  archivePrefix = {arXiv},
  primaryClass = {gr-qc}
}

@article{Estelles2022IMRPhenomTPHM,
  author = {Estell{\'e}s, H{\'e}ctor and Colleoni, Marta and
    Garc{\'i}a-Quir{\'o}s, Cecilio and Husa, Sascha and Keitel, David and
    Mateu-Lucena, Maite and Planas, Maria de Lluc and Ramos-Buades, Antoni},
  title = {New twists in compact binary waveform modeling: A fast time-domain
    model for precession},
  journal = {Phys. Rev. D},
  volume = {105},
  pages = {084040},
  year = {2022},
  doi = {10.1103/PhysRevD.105.084040},
  eprint = {2105.05872},
  archivePrefix = {arXiv},
  primaryClass = {gr-qc}
}

@article{LittenbergCornish2023GLASS,
  author = {Littenberg, Tyson B. and Cornish, Neil J.},
  title = {Prototype global analysis of {LISA} data with multiple source types},
  journal = {Phys. Rev. D},
  volume = {107},
  pages = {063004},
  year = {2023},
  doi = {10.1103/PhysRevD.107.063004},
  eprint = {2301.03673},
  archivePrefix = {arXiv},
  primaryClass = {gr-qc}
}

@article{Katz2025Erebor,
  author = {Katz, Michael L. and Karnesis, Nikolaos and Korsakova, Natalia and
    Gair, Jonathan R. and Stergioulas, Nikolaos},
  title = {Efficient {GPU}-accelerated multisource global fit pipeline for
    {LISA} data analysis},
  journal = {Phys. Rev. D},
  volume = {111},
  pages = {024060},
  year = {2025},
  doi = {10.1103/PhysRevD.111.024060},
  eprint = {2405.04690},
  archivePrefix = {arXiv},
  primaryClass = {gr-qc}
}

@article{Deng2025ModularGlobalFit,
  author = {Deng, Senwen and Babak, Stanislav and Le Jeune, Maude and Marsat,
    Sylvain and Plagnol, {\'E}ric and Sartirana, Andrea},
  title = {Modular global-fit pipeline for {LISA} data analysis},
  journal = {Phys. Rev. D},
  volume = {111},
  pages = {103014},
  year = {2025},
  doi = {10.1103/PhysRevD.111.103014},
  eprint = {2501.10277},
  archivePrefix = {arXiv},
  primaryClass = {gr-qc}
}

@article{Vallisneri2005GeometricTDI,
  author = {Vallisneri, Michele},
  title = {Geometric time delay interferometry},
  journal = {Phys. Rev. D},
  volume = {72},
  pages = {042003},
  year = {2005},
  doi = {10.1103/PhysRevD.72.042003},
  eprint = {gr-qc/0504145},
  archivePrefix = {arXiv},
  note = {Erratum: Phys. Rev. D 76, 109903 (2007)}
}

@article{pycwb,
    title = {{PycWB}: A user-friendly, Modular, and python-based framework for gravitational wave unmodelled search},
    journal = {SoftwareX},
    volume = {26},
    pages = {101639},
    year = {2024},
    issn = {2352-7110},
    doi = {10.1016/j.softx.2024.101639},
    url = {https://www.sciencedirect.com/science/article/pii/S2352711024000104},
    author = {Yumeng Xu and Shubhanshu Tiwari and Marco Drago}
}

@article{Garg2024LISASystematics,
  author = {Garg, Mudit and Sberna, Laura and Speri, Lorenzo
            and Duque, Francisco and Gair, Jonathan},
  title = {Systematics in tests of general relativity using
           {LISA} massive black hole binaries},
  journal = {Monthly Notices of the Royal Astronomical Society},
  volume = {535},
  number = {4},
  pages = {3283--3292},
  year = {2024},
  doi = {10.1093/mnras/stae2605},
  eprint = {2410.02910},
  archivePrefix = {arXiv},
  primaryClass = {astro-ph.GA}
}

@article{Deng2025FastReconstruction,
  author = {Deng, Senwen and Babak, Stanislav and Marsat, Sylvain},
  title = {Fast detection and reconstruction of merging massive black hole binary signals},
  journal = {Phys. Rev. D},
  volume = {112},
  pages = {043010},
  year = {2025},
  doi = {10.1103/jyr7-fcgp},
  eprint = {2504.11322},
  archivePrefix = {arXiv},
  primaryClass = {gr-qc}
}

@article{Cornish2020TimeFrequency,
  author = {Cornish, Neil J.},
  title = {Time-frequency analysis of gravitational wave data},
  journal = {Phys. Rev. D},
  volume = {102},
  pages = {124038},
  year = {2020},
  doi = {10.1103/PhysRevD.102.124038},
  eprint = {2009.00043},
  archivePrefix = {arXiv},
  primaryClass = {gr-qc}
}

@article{Cornish2022LowLatency,
  author = {Cornish, Neil J.},
  title = {Low latency detection of massive black hole binaries},
  journal = {Phys. Rev. D},
  volume = {105},
  pages = {044007},
  year = {2022},
  doi = {10.1103/PhysRevD.105.044007},
  eprint = {2110.06238},
  archivePrefix = {arXiv},
  primaryClass = {gr-qc}
}

@article{Cornish2025WaveletCovariance,
  author = {Cornish, Neil J.},
  title = {Nonstationary noise in gravitational wave analyses: The wavelet domain noise covariance matrix},
  journal = {Phys. Rev. D},
  volume = {114},
  pages = {044035},
  year = {2026},
  doi = {10.1103/n5rl-fpqw},
  eprint = {2511.10632},
  archivePrefix = {arXiv},
  primaryClass = {gr-qc}
}

@article{Cornish2026DynamicNoise,
  author = {Cornish, Neil J.},
  title = {Modeling nonstationary noise: Applications in gravitational wave astronomy},
  journal = {Phys. Rev. D},
  volume = {114},
  pages = {064054},
  year = {2026},
  doi = {10.1103/vfvw-rhvf},
  eprint = {2607.13168},
  archivePrefix = {arXiv},
  primaryClass = {gr-qc}
}

@misc{Cornish2026FastLISATemplates,
  author = {Cornish, Neil J.},
  title = {Time-frequency analysis for {LISA}: Fast waveform templates},
  year = {2026},
  eprint = {2609.28316},
  archivePrefix = {arXiv},
  primaryClass = {gr-qc},
  url = {https://arxiv.org/abs/2609.28316}
}

@article{DigmanCornish2022GalacticBackground,
  author = {Digman, Matthew C. and Cornish, Neil J.},
  title = {{LISA} Gravitational Wave Sources in a Time-varying Galactic Stochastic Background},
  journal = {Astrophys. J.},
  volume = {940},
  pages = {10},
  year = {2022},
  doi = {10.3847/1538-4357/ac9139},
  eprint = {2206.14813},
  archivePrefix = {arXiv},
  primaryClass = {astro-ph.IM},
  url = {https://arxiv.org/abs/2206.14813}
}

@article{DigmanCornish2023StellarBH,
  author = {Digman, Matthew C. and Cornish, Neil J.},
  title = {Parameter estimation for stellar-origin black hole mergers in {LISA}},
  journal = {Phys. Rev. D},
  volume = {108},
  pages = {023022},
  year = {2023},
  doi = {10.1103/PhysRevD.108.023022},
  eprint = {2212.04600},
  archivePrefix = {arXiv},
  primaryClass = {gr-qc}
}

@misc{Johnson2026WDMCovariance,
  author = {Johnson, Aaron D. and Chatziioannou, Katerina and Rosati, Robert},
  title = {The {WDM} time-frequency transform in gravitational-wave data analysis {II}: Evolutionary spectrum and noise covariance},
  year = {2026},
  eprint = {2610.19002},
  archivePrefix = {arXiv},
  primaryClass = {gr-qc},
  note = {Version 1},
  url = {https://arxiv.org/abs/2610.19002v1}
}

@misc{Johnson2026WDMFormalism,
  author = {Johnson, Aaron D. and Chatziioannou, Katerina and Summers, Jake},
  title = {The {WDM} Time-Frequency Transform in Gravitational-Wave Data Analysis {I}: Formalism},
  year = {2026},
  eprint = {2606.21473},
  archivePrefix = {arXiv},
  primaryClass = {gr-qc},
  note = {Version 2},
  url = {https://arxiv.org/abs/2606.21473v2}
}

@misc{Pearson2026CompactTimeFrequency,
  author = {Pearson, Noah and {\c{C}}al{\i}{\c{s}}kan, Mesut and Bini, Sophie and Cornish, Neil J.},
  title = {A compact time-frequency representation for gravitational-wave data analysis},
  year = {2026},
  eprint = {2609.05601},
  archivePrefix = {arXiv},
  primaryClass = {gr-qc},
  note = {Version 1},
  url = {https://arxiv.org/abs/2609.05601v1}
}

@article{PearsonCornish2026DataGaps,
  author = {Pearson, Noah and Cornish, Neil J.},
  title = {Handling data gaps for the next generation of gravitational-wave observatories},
  journal = {Phys. Rev. D},
  volume = {113},
  pages = {064033},
  year = {2026},
  doi = {10.1103/htg6-w4yy},
  eprint = {2509.05479},
  archivePrefix = {arXiv},
  primaryClass = {gr-qc}
}

@misc{Vajpeyi2026DifferentiableWDM,
  author = {Vajpeyi, Avi and Mentasti, Giorgio and Baghi, Quentin and Burke, Ollie and Speri, Lorenzo},
  title = {An explicit and differentiable {Wilson--Daubechies--Meyer} transform for gravitational-wave data analysis},
  year = {2026},
  eprint = {2606.20269},
  archivePrefix = {arXiv},
  primaryClass = {gr-qc},
  note = {Version 2},
  url = {https://arxiv.org/abs/2606.20269v2}
}

@article{Brown2007ThreeStage,
  author = {Brown, Duncan A. and Crowder, Jeff and Cutler, Curt and Mandel, Ilya and Vallisneri, Michele},
  title = {A three-stage search for supermassive black-hole binaries in {LISA} data},
  journal = {Class. Quantum Grav.},
  volume = {24},
  pages = {S595},
  year = {2007},
  doi = {10.1088/0264-9381/24/19/S22},
  eprint = {0704.2447},
  archivePrefix = {arXiv},
  primaryClass = {gr-qc}
}

@article{Nobili2026FastPremerger,
  author = {Nobili, Francesco and Bellotti, Malvina and Buscicchio, Riccardo and Dotti, Massimo and Lupi, Alessandro},
  title = {Fast premerger detection of massive black-hole binaries in {LISA} based on time-frequency excess power},
  journal = {Phys. Rev. D},
  volume = {114},
  pages = {063001},
  year = {2026},
  doi = {10.1103/l85g-ds5n},
  eprint = {2602.16792},
  archivePrefix = {arXiv},
  primaryClass = {astro-ph.IM}
}

@article{Bandopadhyay2026GlobalTF,
  author = {Bandopadhyay, Diganta and Chapman-Bird, Christian E. A. and Vecchio, Alberto},
  title = {Global time-frequency search for stellar-mass binary black holes in {LISA}},
  journal = {Phys. Rev. D},
  volume = {114},
  pages = {064015},
  year = {2026},
  doi = {10.1103/339z-w985},
  eprint = {2510.19047},
  archivePrefix = {arXiv},
  primaryClass = {gr-qc}
}

@article{Speri2026SingleHarmonic,
  author = {Speri, Lorenzo and Tenorio, Rodrigo and Chapman-Bird, Christian and Gerosa, Davide},
  title = {Single-harmonic search for extreme mass-ratio inspirals},
  journal = {Phys. Rev. D},
  volume = {113},
  pages = {024061},
  year = {2026},
  doi = {10.1103/dh3j-ksfl},
  eprint = {2510.20891},
  archivePrefix = {arXiv},
  primaryClass = {gr-qc}
}

@article{TenorioGerosa2025SFT,
  author = {Tenorio, Rodrigo and Gerosa, Davide},
  title = {Scalable data-analysis framework for long-duration gravitational waves from compact binaries using short {Fourier} transforms},
  journal = {Phys. Rev. D},
  volume = {111},
  pages = {104044},
  year = {2025},
  doi = {10.1103/PhysRevD.111.104044},
  eprint = {2502.11823},
  archivePrefix = {arXiv},
  primaryClass = {gr-qc}
}

@article{GWTC2,
  author        = {Abbott, R. and others},
  collaboration = {LIGO Scientific Collaboration and Virgo Collaboration},
  title         = {{GWTC-2}: Compact Binary Coalescences Observed by
                   {LIGO} and {Virgo} during the First Half of the
                   Third Observing Run},
  journal       = {Phys. Rev. X},
  volume        = {11},
  pages         = {021053},
  year          = {2021},
  doi           = {10.1103/PhysRevX.11.021053},
  eprint        = {2010.14527},
  archivePrefix = {arXiv},
  primaryClass  = {gr-qc}
}

@article{GWTC3,
  author        = {Abbott, R. and others},
  collaboration = {LIGO Scientific Collaboration, Virgo Collaboration,
                   and KAGRA Collaboration},
  title         = {{GWTC-3}: Compact Binary Coalescences Observed by
                   {LIGO} and {Virgo} during the Second Part of the
                   Third Observing Run},
  journal       = {Phys. Rev. X},
  volume        = {13},
  pages         = {041039},
  year          = {2023},
  doi           = {10.1103/PhysRevX.13.041039},
  eprint        = {2111.03606},
  archivePrefix = {arXiv},
  primaryClass  = {gr-qc}
}

@article{GWTC4,
  author        = {Abac, A. G. and others},
  collaboration = {LIGO Scientific Collaboration, Virgo Collaboration,
                   and KAGRA Collaboration},
  title         = {{GWTC-4.0}: Updating the Gravitational-wave Transient
                   Catalog with Observations from the First Part of the
                   Fourth {LIGO--Virgo--KAGRA} Observing Run},
  journal       = {Astrophys. J. Lett.},
  volume        = {1004},
  pages         = {L22},
  year          = {2026},
  doi           = {10.3847/2041-8213/ae2c74},
  eprint        = {2508.18082},
  archivePrefix = {arXiv},
  primaryClass  = {gr-qc}
}

@article{Cornish2021BayesWave,
  author        = {Cornish, Neil J. and Littenberg, Tyson B. and
                   B{\'e}csy, Bence and Chatziioannou, Katerina and
                   Clark, James A. and Ghonge, Sudarshan and
                   Millhouse, Margaret},
  title         = {{BayesWave} analysis pipeline in the era of
                   gravitational wave observations},
  journal       = {Phys. Rev. D},
  volume        = {103},
  pages         = {044006},
  year          = {2021},
  doi           = {10.1103/PhysRevD.103.044006},
  eprint        = {2011.09494},
  archivePrefix = {arXiv},
  primaryClass  = {gr-qc}
}

\end{document}